\documentclass[twocolumn,showpacs,showkeys,amsmath,amssymb,prd,floatfix]{revtex4}
\usepackage{ifthen,graphicx}
\usepackage{bm}

\begin{document}
\title{Is the Mass Scale for Elementary Particles \\
Classically Determined?}

\date{\today}

\author{Peter~R.~Phillips}
\affiliation{Department of Physics, Washington University, St.~Louis,
MO 63130 }
\email{prp@wuphys.wustl.edu}

\begin{abstract}
We investigate whether a mass scale for elementary particles can be derived
from interactions of particles with the distant matter in the Universe, the
mechanism of the interaction being the classical vector potential,
propagating in a space of negative curvature. A possible context for such a
mass scale is conformal gravity. This theory may prove to be renormalizable,
since all coupling constants are dimensionless; conversely, however, there
is no coupling constant analogous to the conventional $G$ to provide a
starting point for a mass scale calculation. We obtain the equations for
propagation of the vector potential of a charged particle moving in a plasma
in a curved space. We then show that distant matter will contribute to
$\langle A^{\mu} A_{\mu} \rangle$, and that this non-thermal part will
eventually dominate the ordinary thermal part. At this point a symmetry
breaking transition of the Coleman-Weinberg type is possible, and particle
masses can be generated with $m^2 \approx \langle A^{\mu} A_{\mu} \rangle$.


\end{abstract}


\maketitle


\section{\label{sec:intro}INTRODUCTION}

In this paper we will try to implement a possibility suggested in a
well-known text on quantum field theory \cite{pesk1}. After surveying the
difficulties faced by current theories of the mass scale of elementary
particles, the authors write: ``\ldots it may be that the overall scale of
energy-momentum is genuinely ambiguous and is set by a cosmological boundary
condition.'' The difficulties the authors are concerned about begin to
appear in the usual description of the origin of the mass scale:
\begin{enumerate}
	\item The Einstein theory of gravity, based on an action that is linear
	in the curvature tensor, is the correct one.
	\item The coupling constants of the fundamental interactions are all
	dimensionless except for the gravitational constant, $G$. As a result,
	all interactions are renormalizable except for gravity.
	\item $G$ can be combined with other constants to define a mass, the
	Planck mass, $m_{\mathrm{planck}}$.
	\item Because gravity is not renormalizable, the masses of particles
	will be dominated by high-energy gravitational self-interactions, and
	the masses should be of the same order as the Planck mass.
\end{enumerate}

This brings us at once to the ``hierarchy problem'': why is the ratio
$m_{\mathrm{proton}}/m_{\mathrm{planck}}$ so small, about $10^{-19}$?

In this paper we present a very different model of the origin of the mass
scale, with these main features:
\begin{enumerate}
	\item A gravitational theory that is renormalizable, and has
	dimensionless coupling constants, will ultimately come to be accepted
	in place of the Einstein theory. A possible example of such a theory is
	conformal gravity, which is based on an action that is quadratic
	in the curvature tensor (see \cite{mann6}, and other papers cited
	there). It is too early to say whether conformal gravity will be able
	to describe recent cosmological observations, but we will assume that
	whatever theory is ultimately accepted will have the basic
	characteristics introduced here. In most of this paper the precise
	form of the gravitational field equations will be irrelevant. We will
	only need these equations in appendix \ref{app:Mannheim}, where we
	consider Mannheim's model \cite{mann6} of conformal gravity.
	\item In a renormalizable theory the mass scale may not be generated
	by high-energy virtual processes at all, and we suggest that it
	actually arises from interaction with distant matter in the Universe.
	\item The agent of this interaction is the familiar classical vector
	potential of the electromagnetic field, propagating in a space of
	negative curvature.
	\item The influence of distant matter rises steadily from the birth of
	the Universe until a symmetry-breaking transition takes place, possibly
	of the Coleman-Weinberg type \cite{cole1}, when masses as we know them
	appear. We use the letters CW, in text and subscripts, to refer to this
	transition.
\end{enumerate}

We start from the simple, and at first sight pointless,
observation that the kinetic-energy term of the Klein-Gordon equation for
a charged scalar field in Minkowski space,
$\phi^{*} ({\mathbf p} - q{\mathbf A})^2 \phi$, when expanded, gives the
term $\phi^{*} q^2 {\mathbf A}^2 \phi$. This has the same sign as the mass
term $\phi^{*} m^2 \phi$, suggesting that in some circumstances, such as the
radiation era of the early Universe, $\langle q^2 {\mathbf A}^2 \rangle$
might play the role of $m^2$. In the familiar world of of Minkowski space
this will not happen, because $\langle {\mathbf A}^2 \rangle$ depends simply
on the local temperature. When pulses propagate in a curved space, however,
the potentials (though not the fields) leave a tail, as will be shown in
detail in section \ref{sec:propvec}. As a result, distant matter will
generate a non-thermal component in $\langle {\mathbf A}^2 \rangle$.

Our model will use the preferred geometry for conformal cosmology
\cite{mann2}, a Friedmann-Robertson-Walker (FRW) Universe with negative
curvature. The current consensus among cosmologists is that space is flat,
at least at the present time. But we must remember that the common
inference that space is flat rests on the usual assumptions about gravity
and the generation of mass. In this paper these assumptions are in
abeyance, so the curvature of space remains an open question.

The early Universe is filled with radiation-dominated plasma, in which each
charged particle is accompanied by a screening cloud. It is by no means
obvious how electromagnetic interactions could extend over cosmological
distances, rather than declining exponentially over distances of the order
of a Debye length. In section \ref{sec:screen} we explain how this happens,
and describe our model of the sources of these long-range fields and
potentials.

In subsequent sections we study the propagation of electromagnetic waves in
curved space. Here we will not need any gravitational field equations,
but will simply use the well-known equations for propagation of a classical
electromagnetic field \cite{mtw}. However, we are assuming that the vector
potential can in some circumstances act as a direct agent, rather than
through its derivatives, so the gauge cannot be freely chosen and we will
have to pay careful attention to the gauge condition. We discuss
gauge invariance and the gauge condition later, in section \ref{sec:dipot}.

In sections \ref{sec:propvec} and \ref{sec:mass_1} we give the simplest
derivation of the mass scale, assuming that the long-range fields from the
plasma are not thermalized, even when propagating in a curved space.

Sections \ref{sec:introcond} through \ref{sec:mass_2} can be omitted on
first reading. Here we assume the long-range fields are slowly thermalized,
and verify that even in this situation the mass scale will be established.

Section \ref{sec:conclude} gives our conclusions.

\section{\label{sec:notation}NOTATION}

Units are chosen so that $c = \hbar = 1$. Our sign conventions are
those of Weinberg \cite{wein2}, so the metric signature is $( - + + + )$
and $g = -{\mathrm{Det}}\,(g_{\mu\nu})$.

For the early Universe, assumed spatially homogeneous, we use a FRW metric,
with coordinates $t$, $\chi$, $\theta$, $\phi$,
labeled 0, 1, 2, 3:
\ifthenelse {\lengthtest{\baselineskip > 16pt}}	
{
	\begin{eqnarray}
	\mathrm{d}s^2 & = & -\mathrm{d}t^2 +
	R^2 (t) \left( \mathrm{d}\chi^2 + \sinh^2 \chi \, \mathrm{d}\theta^2
	+ \sinh^2 \chi \sin^2 \theta \, \mathrm{d}\phi^2 \right) ,
	\label{eq:ds1} \\
	g_{\mu\nu} & = & {\mathrm{diag}} \left( -1, R^2 (t),
	R^2 (t) \sinh^2 \chi, R^2 (t) \sinh^2 \chi \sin^2 \theta \right) ,
	\label{eq:gmetric1} \\
	\sqrt{g} & = & R^3 (t) \sinh^2 \chi \sin \theta  . \label{eq:gdet1} 
	\end{eqnarray}
} {	
	\begin{eqnarray}
	\mathrm{d}s^2 & = & -\mathrm{d}t^2 +
	R^2 (t) \left( \mathrm{d}\chi^2 + \sinh^2 \chi \, \mathrm{d}\theta^2
	\right. \nonumber \\
	  & & {} \left. + \sinh^2 \chi \sin^2 \theta \, \mathrm{d}\phi^2 \right) ,
	\label{eq:ds1} \\
	g_{\mu\nu} & = & {\mathrm{diag}} \left( -1, R^2 (t),
	R^2 (t) \sinh^2 \chi, \right. \nonumber \\
	  & &  \left. R^2 (t) \sinh^2 \chi \sin^2 \theta
	\right) , \label{eq:gmetric1} \\
	\sqrt{g} & = & R^3 (t) \sinh^2 \chi \sin \theta  . \label{eq:gdet1} 
	\end{eqnarray}
}	
The expansion parameter, $R(t)$, has the dimension of length, and can be
thought of as the radius of the Universe at time $t$.

The conformal time, $\eta$, is related to the ordinary time $t$ by
$\mathrm{d}\eta = \mathrm{d}t/R(t)$. In terms of $\eta$, $\chi$, $\theta$,
$\phi$ (all dimensionless) we have:
\ifthenelse {\lengthtest{\baselineskip > 16pt}}	
{
	\begin{eqnarray}
	\mathrm{d}s^2 & = & R^2 (\eta)\left(-\mathrm{d} \eta^2 +
	\mathrm{d}\chi^2 + \sinh^2 \chi \, \mathrm{d}\theta^2
	+ \sinh^2 \chi \sin^2 \theta \, \mathrm{d}\phi^2 \right) ,
	\label{eq:ds2} \\
	g_{\mu\nu} & = & R^2 (\eta){\mathrm{diag}} \left( -1, 1, \sinh^2 \chi,
	\sinh^2 \chi \sin^2 \theta \right) , \label{eq:gmetric2} \\
	\sqrt{g} & = & R^4 (\eta) \sinh^2 \chi \sin \theta . \label{eq:gdet2} 
	\end{eqnarray}
} {	
	\begin{eqnarray}
	\mathrm{d}s^2 & = & R^2 (\eta)\left(-\mathrm{d} \eta^2 +
	\mathrm{d}\chi^2 + \sinh^2 \chi \, \mathrm{d}\theta^2 \right. \nonumber \\
	  & & {} \left. + \sinh^2 \chi \sin^2 \theta \, \mathrm{d}\phi^2 \right) ,
	\label{eq:ds2} \\
	g_{\mu\nu} & = & R^2 (\eta){\mathrm{diag}} \left( -1, 1, \sinh^2 \chi,
	\sinh^2 \chi \sin^2 \theta \right) , \label{eq:gmetric2} \\
	\sqrt{g} & = & R^4 (\eta) \sinh^2 \chi \sin \theta . \label{eq:gdet2} 
	\end{eqnarray}
}	

Maxwell's equations in free space are:
\begin{eqnarray}
\left( \sqrt{g} F^{\alpha \beta} \right) _{,\beta} & = & 0 ,
\label{eq:max1} \\
F_{\alpha \beta , \gamma} + F_{\beta \gamma , \alpha} 
+ F_{\gamma \alpha , \beta} & = & 0 . \label{eq:max2}
\end{eqnarray}

Potentials and the gauge condition will be introduced later.

\section{\label{sec:model}THE MODEL}

We will assume the Universe has the form of a bubble that condenses out of
a metastable exterior vacuum \cite{cole2,buch1}. The surface of the bubble
expands at the speed of light, and the interior of the bubble forms a FRW
space with negative curvature. We are not concerned with inflation, but will
simply assume that the material content of the Universe appears as
radiation at the surface of the bubble.

We restrict the material content of our model to the following:
\begin{enumerate}
	\item A charged, massless, scalar field, $\phi$, with conformal weight
	$-1$.
	\item The usual electromagnetic fields and potentials.
	\item In appendix \ref{app:Mannheim} we implicitly assume, following
	Mannheim \cite{mann6}, an additional neutral scalar field, $S$, which
	from the beginning has a very large, constant value, $S_0$, but will
	play no further role.
\end{enumerate}

\section{\label{sec:screen}THE SCREENING CLOUD}

We will be concerned with the propagation of the vector potential generated
by moving charges in the plasma of the early Universe. The charges will have
a thermal mass due to their interaction with the photons, and the plasma
will be relativistic. For simplicity, however, we will carry over some
results from non-relativistic plasmas.

A stationary charged particle in the plasma will be surrounded by a
screening cloud that reduces the field exponentially on a scale known as
the Debye length (\cite{stur}, chapter 2). Surprisingly, however, starting
in the 1950's several authors \cite{neuf,tapp1,mont1} discovered that
a test charge \textit{moving uniformly} is not exponentially screened, but
generates the field of a quadrupole in a collisionless plasma. The
restriction to a collisionless plasma was lifted in later papers 
\cite{yu,schr}, where it was shown that the far field of a moving
test charge is that of a dipole with strength of order $q\tau_t V$, where
$q$ is the charge, $\tau_t$ the collision time as measured by $t$, and
$V$ the velocity.
 
A charge that is part of the plasma, rather than being a test charge
constrained to move in a certain way, will itself experience collisions. We
can picture such a particle as describing Brownian motion, with its
screening cloud sometimes closer, sometimes farther, but never fully
established. We will model the most important aspect of a single step of
this process by setting up a local set of Cartesian coordinates and placing
a stationary screening charge $-q$ at the origin. A charge $q$ is initially
at rest on the $z$ axis at $z = -V \tau_t$, then begins to move with uniform
velocity $V$ along the $z$ axis, starting at $t = -\tau_t$ and ending at
$t = \tau_t$, when it again comes to rest. Note that this choice of origin
for $t$ differs from that of the cosmic time used in section
\ref{sec:model}; $\eta$, defined below, will differ in a similar way. We
reconcile these differences later, in section \ref{sec:mass_1}.

We are now going to take the result for a uniformly moving test charge and
use it to calculate the field around our model charge that is subject
to Brownian motion. We justify this step as follows. Imagine that the test
charge, instead of moving uniformly from an indefinite time in the past,
is actually stationary until time $t=0$, and then starts moving uniformly.
There will be transient fields, but after a few collision times the final
field of a moving dipole will be established. During the transient period,
the test charge will be moving away from its screening charge, which only
gradually picks up speed and trails along behind. This is similar to the
motion of the charge in our model. Since the transient fields form a
bridge between the initial and final fields they must have a similar form
to the field of the test charge, and extend comparably far.

During the period that the charge is moving (the \textit{pulse}), the dipole
moment will be $\overline{D} (t) = qVt$. Before and after the pulse the
dipole moment remains at a constant value, but this is of little interest
because a constant dipole moment generates no magnetic field or vector
potential. It is well known that Maxwell's equations separate in conformal
time, so we will write the dipole moment in terms of $\eta$. Define
$\tau = \tau_t/R_{\mathrm{s}}(t)$ (dimensionless), so that the pulse extends
from $\eta = -\tau$ to $\eta = \tau$. The subscript `s' on $R_{\mathrm{s}}$
indicates the source time. $R_{\mathrm{s}}(t)$ can be treated as constant
during the pulse, and the dipole moment can be expressed as a Fourier
integral:
\begin{eqnarray}
\overline{D} (\eta) & = & \frac{1}{2\pi} \int D(n) \exp (-in\eta)
\, \mathrm{d}n , \label{eq:dipfour} \\ 
D(n) & = & \frac{2iqVR_{\mathrm{s}} \sin (n\tau)}{n^2} . 
\label{eq:dipstrength}
\end{eqnarray}

We estimate the value of $\tau_t$ in section \ref{sec:mass_1}. The $z$ axis of
the local system used in this section will be parallel to the polar axis of
the polar coordinates used in the bulk of the paper.

\section{\label{sec:dipfield}DIPOLE FIELDS}

We will need only the three fields $F_{10}$, $F_{20}$ and $F_{12}$.
Since Maxwell's equations separate in conformal time the elementary
solutions can be written
\begin{eqnarray}
F_{10} & = & f_{10} (\chi, \theta, \phi) \exp(-in\eta) ,
\label{eq:F10def} \\
F_{20} & = & f_{20} (\chi, \theta, \phi) \exp(-in\eta) ,
\label{eq:F20def} \\
F_{12} & = & f_{12} (\chi, \theta, \phi) \exp(-in\eta) .
\label{eq:F12def}
\end{eqnarray}

For dipole fields we try the following forms:
\begin{eqnarray}
f_{10} & = & f_1 (\chi) P_1 (\theta) , \label{eq:f10} \\
f_{20} & = & f_2 (\chi) N_1 (\theta) , \label{eq:f20} \\
f_{12} & = & f_3 (\chi) N_1 (\theta) , \label{eq:f12}
\end{eqnarray}
with angular functions $P_1 (\theta) = \cos \theta$ and 
$N_1 (\theta) \equiv \mathrm{d} P_1 (\theta)/\mathrm{d} \theta
= -\sin \theta$.

Maxwell's equations now give (with a prime meaning
$\mathrm{d}/\mathrm{d}\chi$):
\begin{eqnarray}
in \sinh^2 \chi f_1 - 2 f_3 & = & 0 , \label{eq:dip1} \\
in f_2 - f_3^{\,\prime} & = & 0 , \label{eq:dip2} \\
f_1 - f_2^{\,\prime} + in f_3 & = & 0 . \label{eq:dip3} 
\end{eqnarray}

\subsection{\label{subsec:magnetic}Dipole magnetic field}

From (\ref{eq:dip1}), (\ref{eq:dip2}) and (\ref{eq:dip3}) we obtain the
equation for $f_3$ alone
\begin{equation}
\frac{\mathrm{d}^2 f_3}{\mathrm{d} \chi^2} +
\left(n^2 - \frac{2}{\sinh^2 \chi} \right) f_3 = 0 . \label{eq:f3}
\end{equation}

This is the analog of equation (16) of Mashhoon \cite{mash1}, which he
derived for a space of positive curvature. We solve (\ref{eq:f3}) by first
expressing the solution in terms of hypergeometric functions
\cite{absteg}. We then show that these particular
hypergeometric functions can themselves be expressed in closed form in
terms of simpler functions. The indicial equation is $p(p-1) - 2 = 0$, so
$p=2$ or $p=-1$. Write the regular and singular solutions as
\begin{eqnarray}
f_{3,{\mathrm{reg}}} & = & \sinh^2 \chi F_{3,{\mathrm{reg}}} ,
\label{eq:F_reg} \\
f_{3,{\mathrm{sing}}} & = & \sinh^{-1} \chi F_{3,{\mathrm{sing}}} .
\label{eq:F_sing}
\end{eqnarray}
It is then straightforward to show that, in terms of the variable
$\mu = (1 - \cosh \chi)/2$,
\begin{eqnarray}
F_{3,{\mathrm{reg}}} & = &  F (2+in, 2-in; 5/2; \mu) , \label{eq:F_reg2} \\
F_{3,{\mathrm{sing}}} & = &  F (-1+in, -1-in; -1/2; \mu) .
\label{eq:F_sing2}
\end{eqnarray}

Explicit closed forms for these hypergeometric functions are given in
appendix \ref{app:H-G}. $f_3 $ is constructed from that combination of
$f_{3,{\mathrm{reg}}}$ and $f_{3,{\mathrm{sing}}}$ that is proportional to
$\exp (in \chi)$, because when combined with $\exp (-in \eta)$ this
represents outgoing waves:
\begin{equation}
f_3 =  \frac{C_1 (n) \exp (in\chi)}{\sinh \chi}
\left( \cosh \chi - in \sinh \chi \right) , \label{eq:f3out}
\end{equation}
where $C_1 (n)$ is a normalizing factor to be determined.

It is convenient at this point to express $f_3$ in terms of
$u=\tanh (\chi/2)$:
\begin{eqnarray}
f_3 & = &  \frac{C_1 (n) \exp (in\chi)}{2 u}
\left( 1 - 2inu + u^2 \right) . \label{eq:f3out2}
\end{eqnarray}

\subsection{\label{subsec:electric}Dipole electric fields}

From (\ref{eq:dip1}) and (\ref{eq:f3out}) we derive the
equation for the radial electric dipole field:
\begin{eqnarray}
f_1 & = & \frac{-i C_1 (n) \exp(in \chi ) (1-u^2)^2}{4nu^3}
\left( 1 - 2inu + u^2 \right) . \qquad \label{eq:f1out}
\end{eqnarray}

Similarly, from (\ref{eq:dip2}) and (\ref{eq:f3out2}) we derive the
equation for the transverse electric field:
\ifthenelse {\lengthtest{\baselineskip > 16pt}}	
{
	\begin{equation}
	f_2 = \frac{iC_1 (n) \exp(in\chi)}{4nu^2}
	\left[ (1-u^2)^2 - 2inu(1+u^2) - 4n^2u^2 \right] .  \label{eq:f2out}
	\end{equation}
} {	
	\begin{eqnarray}
	f_2 & = & \frac{iC_1 (n) \exp(in\chi)}{4nu^2} \nonumber \\
	  & & \times \left[ (1-u^2)^2 - 2inu(1+u^2) - 4n^2u^2 \right] .
	\label{eq:f2out}
	\end{eqnarray}
}	

\subsection{\label{subsec:normfield}Normalizing the fields}

From (\ref{eq:dipfour}) and (\ref{eq:dipstrength}) we can get the
near-field expression for the radial component of $E$, and by comparison
with (\ref{eq:f1out}) arrive at the form of the normalizing factor
$C_1(n)$. On the polar axis, at small distances, the radial ${\mathbf E}$
field at frequency $n$ is
\ifthenelse {\lengthtest{\baselineskip > 16pt}}	
{
	\begin{equation}
	E_r = -\frac{\partial}{\partial r}
	\left(\frac{D(n) \exp(-in\eta)}{r^2} \right)
	= \frac{2D(n) \exp(-in\eta)}{r^3} , \label{eq:Er_near}
	\end{equation}
} {	
	\begin{eqnarray}
	E_r & = & -\frac{\partial}{\partial r}
	\left(\frac{D(n) \exp(-in\eta)}{r^2} \right) \nonumber \\
	  & = & \frac{2D(n) \exp(-in\eta)}{r^3} ,
	\label{eq:Er_near}
	\end{eqnarray}
}	
where $r=R_{\mathrm{s}} \chi$.

In the same limit (small $\chi$), (\ref{eq:f1out}) gives
$f_1 (\chi) = -2i C_1 (n) /(n \chi^3)$, and so, on the axis,
\begin{equation}
F_{10} (\eta, \chi) = -2i C_1 (n) R_{\mathrm{s}}^3 \exp(-i n \eta)/(n r^3) .
\end{equation}

We now transform to a local Minkowski frame with coordinates $t$, $r$:
\ifthenelse {\lengthtest{\baselineskip > 16pt}}	
{
	\begin{equation}
	\overline{F}_{10} (t, r) = \frac{\partial \eta}{\partial t} 
	\frac{\partial \chi}{\partial r} F_{10} (\eta, \chi)
	= -2i C_1 (n) R_{\mathrm{s}} \exp(-i n \eta)/(n r^3) .
	\label{eq:F10bar}
	\end{equation}
} {	
	\begin{eqnarray}
	\overline{F}_{10} (t, r) & = & \frac{\partial \eta}{\partial t} 
	\frac{\partial \chi}{\partial r} F_{10} (\eta, \chi) \nonumber \\
	  & = &  -2i C_1 (n) R_{\mathrm{s}} \exp(-i n \eta)/(n r^3) .
	\label{eq:F10bar}
	\end{eqnarray}
}	

$\overline{F}_{10}$ is just the conventional radial electric field as given
in (\ref{eq:Er_near}), so using (\ref{eq:dipstrength}):
\begin{equation}
C_1 (n) = \frac{-2qV \sin(n\tau)}{n} .
\label{eq:normC1}
\end{equation}

\section{\label{sec:propf12}PROPAGATION OF THE MAGNETIC FIELD}

(\ref{eq:f3out2}), (\ref{eq:normC1}), (\ref{eq:f12}) and (\ref{eq:F12def})
give the Fourier transform of the magnetic field:
\ifthenelse {\lengthtest{\baselineskip > 16pt}}	
{
	\begin{equation}
	F_{12} = \frac{-qV \sin (n \tau) N_1 (\theta) \exp [in(\chi - \eta)]}
	{ n u} \left( 1 - 2inu + u^2 \right) . \label{eq:F12out}
	\end{equation}
} {	
	\begin{eqnarray}
	F_{12} & = & \frac{-qV \sin (n \tau) N_1 (\theta) \exp [in(\chi - \eta)]}
	{ n u} \nonumber \\
	 & & \times \left( 1 - 2inu + u^2 \right) . \label{eq:F12out}
	\end{eqnarray}
}	

\begin{figure}[ht]
\centering
\ifthenelse {\lengthtest{\baselineskip > 16pt}}	
{
	\includegraphics[scale=1.0]{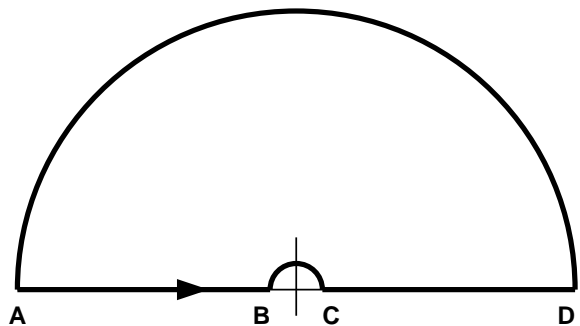}
} {	
	\includegraphics[width=3.0in]{msepfig1.eps}
}	
\caption{\label{fig:contour1} Typical contour for Fourier synthesis of
magnetic and electric fields.  }
\end{figure}

We transform back into $\chi$, $\eta$ space by dividing by $2\pi$ and
integrating over $n$ along the real axis. For $\chi > \eta + \tau$ both
exponentials in $\sin (n \tau)$ allow us to close in the UHP, using the
contour of figure \ref{fig:contour1}. $\sin (n\tau)/n$ is regular at $n=0$,
so we get zero, as required by causality. Similarly, when
$\chi < \eta - \tau$ we can close in the LHP, using a contour that is the
inverse of figure \ref{fig:contour1}, and we again get zero.

For $\eta - \tau < \chi < \eta + \tau$ we must write
$\sin (n \tau) = (\exp(in \tau) - \exp(-in \tau))/2i$, divide the integrand
into two pieces, and close the first integral in the UHP and the second in
the LHP. Combining these two integrals we get
\begin{equation}
F_{12} = \frac{qV \sin \theta (1+u^2)}{2u} . 
\label{eq:F12out2}
\end{equation}

The propagation of $F_{12}$ is shown in figure \ref{fig:HPEPS}. An
unexpected feature of the pulse is that the amplitude does not tend to zero
for large $\chi$, but to a constant asymptotic value. The
\textit{conventional} ${\mathbf H}$ field, does, of course, tend to zero,
in fact exponentially, because in a local Lorentz frame we have
$\overline{F}_{12}=F_{12}/(R_{\mathrm{p}}^2\sinh\chi)$, the subscript `p'
on $R_{\mathrm{p}}$ indicating the point of observation.

\ifthenelse {\lengthtest{\baselineskip > 16pt}}	
{
	\begin{figure}[ht]
	\centering
	\includegraphics[scale=0.9]{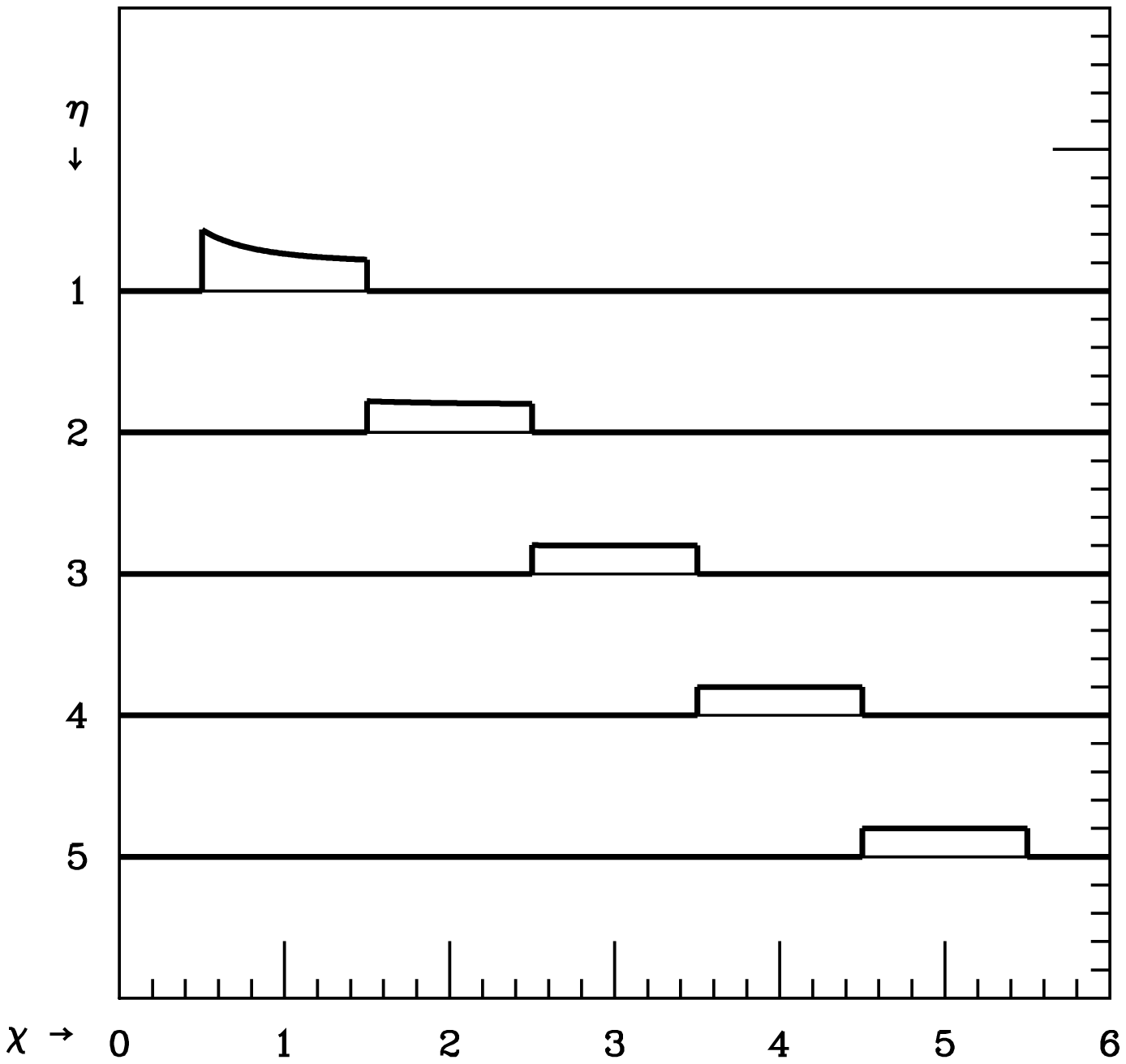}
	\caption{\label{fig:HPEPS} Propagation of the magnetic field, $F_{12}$:
	snapshots of the pulse for various values of conformal time, $\eta$. For
	each $\eta$, the pulse is confined to the range
	$\eta - \tau < \chi < \eta + \tau$; within that range we plot
	$(1+u^2)/u$. To render the pulse visible on the graph, we have
	arbitrarily set $\tau = 0.5$, far larger than it is in practice.  }
	\end{figure}
} {	
}	

\section{\label{sec:propf20}PROPAGATION OF THE ELECTRIC FIELD}

The propagation of the electric field can be displayed in a similar way,
with the contour determined by the boundary conditions for $\eta < -\tau$,
before the pulse begins. Referring to figure \ref{fig:contour1}, we have to
restrict our integral to the portions A--B and C--D, excluding the
semicircle B--C. The integral is interpreted in a principal value sense.
We now find there are electric fields both before and after the main pulse.
This is to be expected, because in our model the source has a static dipole
moment before the pulse begins, and an opposite moment after it has ended.
 
\ifthenelse {\lengthtest{\baselineskip > 16pt}}	
{
} {	
	\begin{figure}[t]
	\centering
	\includegraphics[width=3.375in]{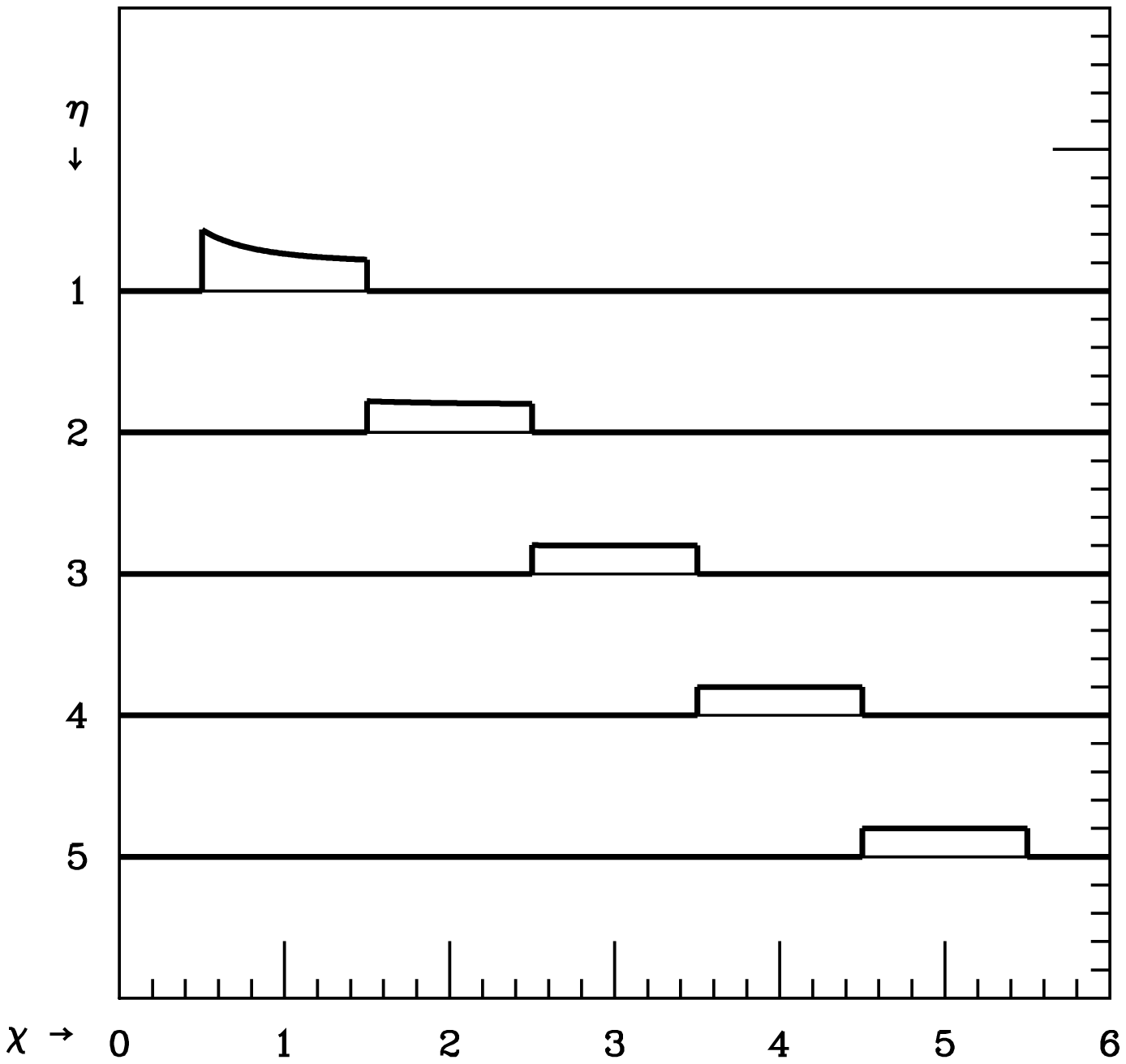}
	\caption{\label{fig:HPEPS} Propagation of the magnetic field, $F_{12}$:
	snapshots of the pulse for various values of conformal time, $\eta$. For
	each $\eta$, the pulse is confined to the range
	$\eta - \tau < \chi < \eta + \tau$; within that range we plot
	$(1+u^2)/u$. To render the pulse visible on the graph, we have
	arbitrarily set $\tau = 0.5$, far larger than it is in practice.  }
	\end{figure}
}	

\section{\label{sec:dipot}DIPOLE POTENTIALS}

The potentials are defined by the following equations:
\begin{eqnarray}
A_0 & = & h_0 (\chi) P_1 (\theta) \exp (-in \eta) ,
\label{eq:Adef0} \\
A_1 & = & h_1 (\chi) P_1 (\theta) \exp (-in \eta) ,
\label{eq:Adef1} \\
A_2 & = & h_2 (\chi) N_1 (\theta) \exp (-in \eta) ,
\label{eq:Adef2} \\
F_{\mu \nu} & = & \frac{\partial A_{\nu}}{\partial x^{\mu}} -
\frac{\partial A_{\mu}}{\partial x^{\nu}} \label{eq:Adef3} .
\end{eqnarray}

From (\ref{eq:Adef3}) we derive:
\begin{eqnarray}
h_2 ^{\,\prime} - h_1 & = & f_3 , \label{eq:heq1} \\
h_0 ^{\,\prime} + in h_1 & = & f_1 , \label{eq:heq2} \\
h_0 + in h_2 & = & f_2 . \label{eq:heq3}
\end{eqnarray}

(\ref{eq:heq1}), (\ref{eq:heq2}) and (\ref{eq:heq3}) are not
independent. If we differentiate (\ref{eq:heq3}), subtract it from
(\ref{eq:heq2}) and use (\ref{eq:dip3}), we obtain (\ref{eq:heq1}).
These three equations therefore do not suffice to define the potentials
uniquely; this is to be expected, because $A_{\mu}$ is only defined up to
a scalar gauge function, $\psi$, so that
$\overline{A}_{\mu} = A_{\mu} + \partial \psi / \partial x^{\mu}$ defines
the same fields as $A_{\mu}$.

To fix the potentials uniquely we need a gauge condition.
The one often suggested is the Lorenz condition \cite{jack2,mtw}:
\begin{equation}
A^{\mu}_{\, ;\mu} = \frac{\partial}{\partial x^{\mu}}
\left(\sqrt{g}\, g^{\mu \nu} A_{\nu} \right) = 0 .  \label{eq:lrnz}
\end{equation}

But this is not acceptable here, because we want a theory that is
conformally invariant, and the Lorenz condition is not (the conformal
weights of $\sqrt{g}$, $g^{\mu\nu}$ and $A_{\mu}$ are $4$, $-2$ and $0$,
respectively). We propose in this paper to use the modified condition
\begin{equation}
\frac{\partial}{\partial x^{\mu}}\left(
\sqrt{g}\, g^{\mu \nu} A_{\nu} \phi^* \phi \right) = 0 .
\label{eq:cnfrm}
\end{equation}

This is conformally invariant since $\phi$ has conformal weight $-1$.
In applying (\ref{eq:cnfrm}) we set $\phi^* \phi$ equal to its expectation
value, which is proportional to $T^2$, and hence to $R^{-2}$. All factors
of $R$ now disappear from (\ref{eq:cnfrm}), and we obtain
\begin{equation}
in \sinh^2 \chi h_0 + \frac{\mathrm{d}}{\mathrm{d} \chi}
\left[ \sinh^2 \chi h_1 \right] - 2 h_2 = 0 .
\label{eq:heq4}
\end{equation}

The potentials are still not defined uniquely. We are permitted to make a
\textit{restricted} gauge transformation, with a gauge function $\psi$
that satisfies
\begin{equation}
\frac{\partial}{\partial x^{\mu}} \left( \sqrt{g} g^{\mu\nu}
\frac{\partial \psi}{\partial x^{\nu}} \right) = 0 .
\label{eq:restrictgauge}
\end{equation}
We will deal with this remaining ambiguity below, in subsection
\ref{subsec:restrict}. 

\subsection{\label{subsec:A0}Scalar potential}

By combining (\ref{eq:dip1}), (\ref{eq:dip2}), (\ref{eq:heq2}),
(\ref{eq:heq3}) and (\ref{eq:heq4}) we can derive an equation for $h_0$
alone:
\begin{equation}
\frac{\mathrm{d}^2 h_0}{\mathrm{d} \chi^2}
+ 2 \coth \chi \frac{\mathrm{d} h_0}{\mathrm{d} \chi}
+ \left( n^2 - \frac{2}{\sinh^2 \chi} \right) h_0 = 0 . \label{eq:h0}
\end{equation}

This is the equation for hyperbolic spherical functions
\cite{buch1,band1}. We will solve it by means of hypergeometric functions,
as for the fields. The indicial equation gives $p = 1$ or $p = -2$, and
the two solutions can be shown to be:
\begin{eqnarray}
h_{\mathrm{0,reg}} & = & \sinh \chi
F(2 + i\alpha, 2 - i\alpha;5/2; \mu) , \label{eq:hreg} \\
h_{\mathrm{0,sing}} & = & \sinh^{-2} \chi
F(-1 + i\alpha, -1 - i\alpha;-1/2; \mu) . \hspace{1em} \label{eq:hsing} 
\end{eqnarray}
where $\mu = (1 - \cosh \chi )/2$ and $\alpha = \sqrt{n^2 - 1}$. The
hypergeometric functions are the same ones we encountered with $f_3$, but the
powers of $\sinh \chi$ appearing in front of them are different.

$A_0$ will be constructed from that combination of the two solutions that
represents outgoing waves, at least for large $n$, where $\alpha$ is real.
With $u=\tanh (\chi/2)$, as before:
\ifthenelse {\lengthtest{\baselineskip > 16pt}}	
{
	\begin{eqnarray}
	A_0 & \equiv & P_1(\theta) h_0 ( \chi ) \exp (-in\eta ) \nonumber \\
	  & = & C_2 (n) P_1 (\theta) \exp [i(\alpha \chi - n \eta)] 
	\frac{(1 - u^2) \left( 1 - 2i\alpha u + u^2 \right) }{u^2} .
	\label{eq:h0out1}
	\end{eqnarray}
} {	
	\begin{eqnarray}
	A_0 & \equiv & P_1(\theta) h_0 ( \chi ) \exp (-in\eta ) \nonumber \\
	  & = & C_2 (n) P_1 (\theta) \exp [i(\alpha \chi - n \eta)] 
	\nonumber \\
	 & & \times \frac{(1 - u^2) \left( 1 - 2i\alpha u + u^2 \right) }{u^2} ,
	\label{eq:h0out1}
	\end{eqnarray}
}	
where $C_2 (n)$ is a normalization factor that depends on the driving
function. $C_2 (n)$ is, of course, proportional to the normalization factor
$C_1 (n)$ defined in (\ref{eq:normC1}). For small $\chi$,
(\ref{eq:h0out1}), (\ref{eq:heq3}) and (\ref{eq:f2out}) give:
\begin{equation}
C_2 (n) = \frac{iC_1 (n)}{4n} . \label{eq:normC2}
\end{equation}

\subsection{\label{subsec:restrict}Restricted gauge transformations}

For restricted gauge transformations, a suitable gauge function of dipole
form is $\psi = h(\chi) P_1 (\theta) \exp (-in \eta)$, where $h$ satisfies
\begin{equation}
\frac{\partial^2 h}{\partial \chi^2}
+ 2 \coth \chi \frac{\partial h}{\partial \chi}
+ \left( n^2 - \frac{2}{ \sinh^2 \chi } \right) h = 0 . \label{eq:hdef}
\end{equation}
This is the same equation that is satisfied by the scalar potential.
If we include $\psi$ in our definitions of the potentials, $h_0$, for
example, becomes $\overline{h}_0 = h_0 -inh$. But $h_0$ permits no such
addition; it is already completely specified by causality and normalization.

\subsection{\label{subsec:A2}Vector potential}

The interesting component of the vector potential is the transverse one,
$A_{\theta}$, or in our notation $A_2$. (\ref{eq:heq3}) gives
\begin{equation}
h_2 = \frac{-i f_2}{n} + \frac{i h_0}{n} . \label{eq:heq2a}
\end{equation}
This can be used to get $h_2$, and from that $A_2$, since 
$f_2$ and $h_0$ are known from (\ref{eq:f2out}) and (\ref{eq:h0out1}).

Equation (\ref{eq:heq2a}) shows that $A_2$ is the sum of two parts, one
involving $\exp (in\chi)$, the other involving $\exp (i\alpha\chi)$:
\begin{eqnarray}
A_2 & = & \frac{-q V N_1 (\theta) \sin (n \tau) e^{-in \eta}}
{2 n^3 u^2} ({\cal N} + {\cal A}) , \label{eq:A2basic} \\
{\cal N} & = & e^{in \chi} \left[ (1-u^2)^2 - 2inu(1+u^2)
- 4 n^2 u^2 \right] , \hspace{1em} \label{eq:calN} \\ 
{\cal A} & = & e^{i \alpha \chi} (1-u^2) \left[ 
-1 + 2i\alpha u - u^2 \right] . \label{eq:calA}
\end{eqnarray}

\subsection{Gauge invariance in the end}
\label{subsec:gaugeinvar}

We have set up the scalar and vector potentials according to a
``classical prescription'': treat the charges in the plasma as classical
particles, and use the simple criteria of causality and conformal
invariance to fix the potentials uniquely. Once we have done this,
we can represent the charged particles as quantum fields. General gauge
transformations are then permitted, provided the phases of the wave
functions of charged particles are simultaneously transformed in the usual
way.

\section{\label{sec:propvec}PROPAGATION OF THE VECTOR POTENTIAL}

The Fourier synthesis of $A_2$ is carried out, as for the magnetic field,
by dividing by $2\pi$ and then integrating over $n$, along a contour chosen
to respect causality. There is no vector potential before the pulse begins,
so the correct contour is a line parallel to the real axis and slightly
above it. The integral must be taken along the whole path A--D in figure
\ref{fig:contour1}, including the semicircle B--C. This ensures that for
$\chi > \eta + \tau$ the contour can be closed in the UHP and the integral
will be zero. There are two regions of interest,
$\eta - \tau < \chi < \eta + \tau$ (the ``main pulse''), and
$\chi < \eta - \tau$ (the ``tail''). 

\begin{figure}[ht]
\centering
\ifthenelse {\lengthtest{\baselineskip > 16pt}}	
{
	\includegraphics[scale=0.9]{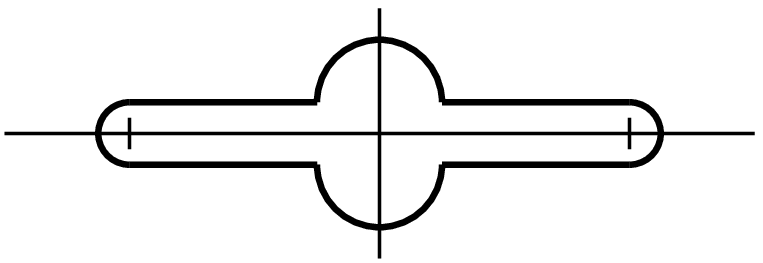}
} {	
	\includegraphics[width=3.375in]{msepfig3.eps}
}	
\caption{\label{fig:contour2} Integration contour for $A_{2}$. The contour
encloses the points $n = -1$ and $n = +1$; it is described in a
clockwise sense. }
\end{figure}

We get a non-zero result only for that part of the integral that involves a
contour that is closed in the LHP. For the part of equation
(\ref{eq:A2basic}) that is proportional to $\exp (in \chi)$ we can shrink
this contour to a small circle about the point $n=0$, and we just have to
find the residue there. For the part that is proportional to
$\exp (i \alpha \chi)$ we have to remember the branch points at $n = \pm 1$,
so our contour can only be shrunk to the form shown in figure
\ref{fig:contour2}.

$A_2$ is the sum of contributions from small circles (or semicircles)
around $n=0$ (the pole terms) and the integrals along the cut from
$n=-1$ to $n=1$. The pole terms have simple analytic expressions, but the
integrals must be evaluated numerically for each ($\eta,\,\chi$) pair.
The integrations are straightforward, and the propagation of $A_2$ is
shown in figure \ref{fig:ATEPS}.

An important difference between figures \ref{fig:HPEPS} and \ref{fig:ATEPS}
is that in the latter the pulse has a non-zero tail for
$\chi < \eta - \tau$. This feature of the propagation of potentials in a
curved space has been noted before \cite{dew,nar1}. For dipole
propagation, as here, the tail rises linearly from $\chi = 0$, and
approaches a constant value for large $\chi$. This asymptotic value is
proportional to $\tau$. In ordinary Minkowski space, which corresponds to
the limit $\chi \rightarrow 0$, ${\mathbf A}$ has no tail.

We note that the following simple function, with $a = 0.735$, gives an
adequate fit for $A_2^2$ for large $\eta$:
\begin{equation}
A_2^2 \approx \left( 2 q V \tau \sin \theta \right)^2 
\tanh^2 (a\chi) . \label{eq:A2fit}
\end{equation}

\ifthenelse {\lengthtest{\baselineskip > 16pt}}	
{
	\begin{figure}[ht]
	\centering
	\includegraphics[scale=0.9]{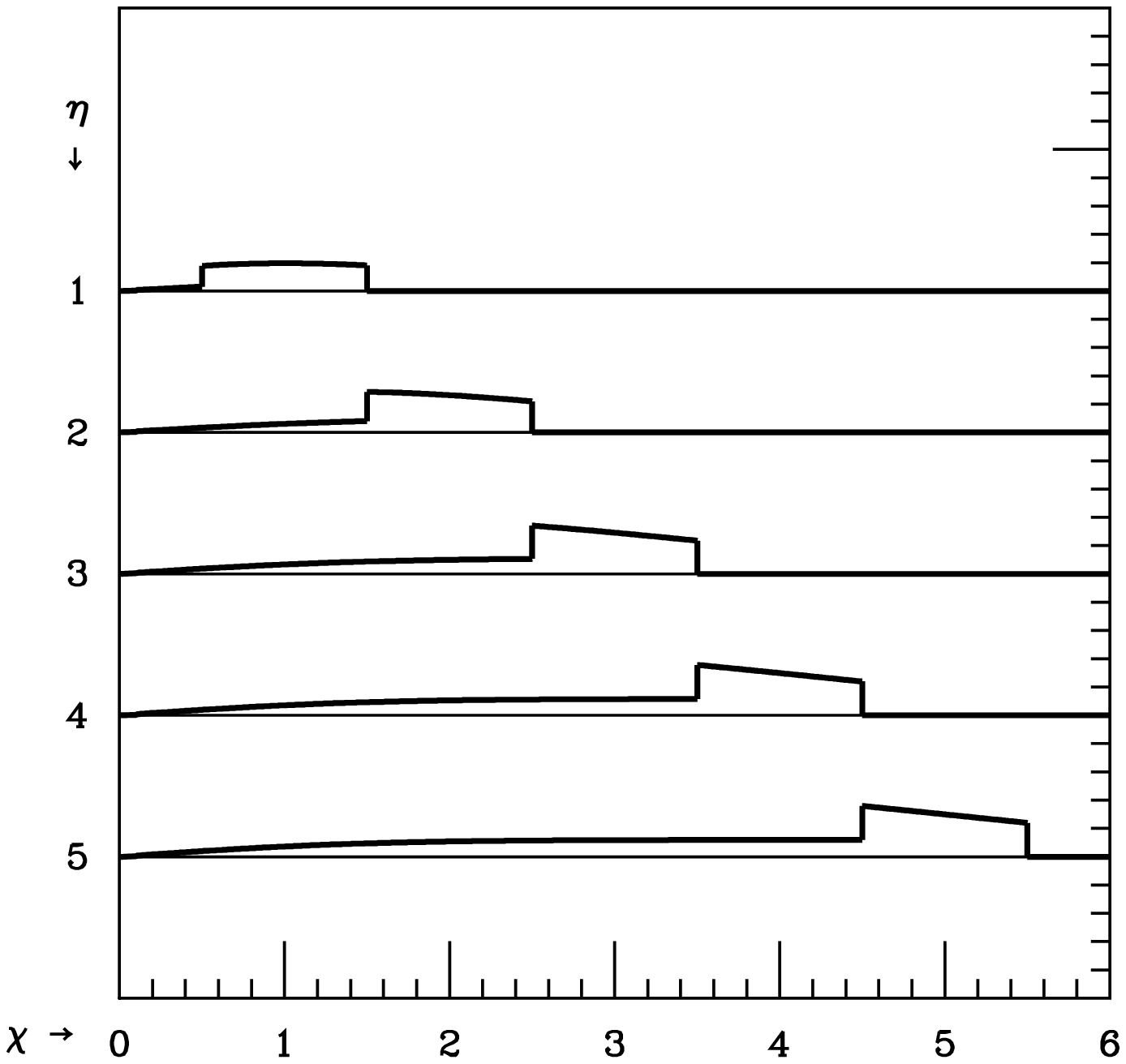}
	\caption{\label{fig:ATEPS} Propagation of the transverse component of
	the vector potential, $A_2$: snapshots of the pulse and the tail for a
	series of values of $\eta$. As with the magnetic field, we set
	$\tau = 0.5$ to make the pulse visible on the graph. }
	\end{figure}
} {	
}	

\section{\label{sec:mass_1}MASS GENERATION: FIRST CALCULATION}

Let us temporarily set aside all questions of gauge invariance, and simply
regard $\langle A^{\mu} A_{\mu} \rangle$ as being composed of a thermal
part, $\langle A^{\mu} A_{\mu} \rangle_{\mathrm{th}}$, which will be of
order $T^2$, and a non-thermal part,
$\langle A^{\mu} A_{\mu} \rangle_{\mathrm{nt}}$.
Moving particles in the distant plasma will generate pulses of $A_{\mu}$,
each of which consists of a ``main pulse'' and a ``tail''. The main pulse
contains electric and magnetic fields, and will contribute to
$\langle A^{\mu} A_{\mu} \rangle_{\mathrm{th}}$. The tail, however, is a
pure gauge potential that generates no fields; it will contribute to
$\langle A^{\mu} A_{\mu} \rangle_{\mathrm{nt}}$.

\ifthenelse {\lengthtest{\baselineskip > 16pt}}	
{
} {	
	\begin{figure}[t]
	\centering
	\includegraphics[width=3.375in]{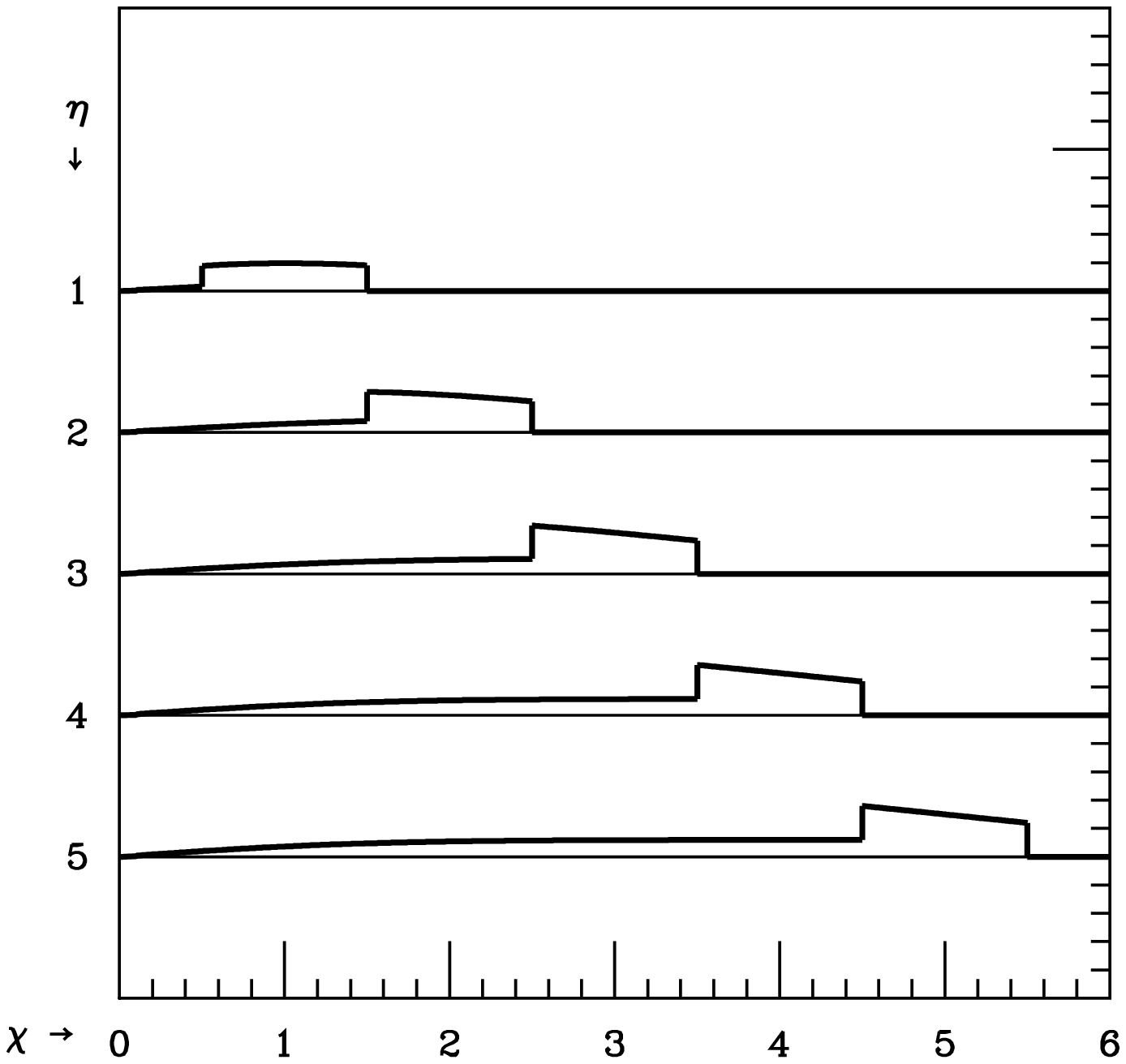}
	\caption{\label{fig:ATEPS} Propagation of the transverse component of
	the vector potential, $A_2$: snapshots of the pulse and the tail for a
	series of values of $\eta$. As with the magnetic field, we set
	$\tau = 0.5$ to make the pulse visible on the graph. }
	\end{figure}
}	

The development of $\langle A^{\mu} A_{\mu} \rangle_{\mathrm{nt}}$ can be
visualized as follows. As each pulse passes the observation point it leaves
a memory in the form of the tail. These tails will not cancel but will add
according to the theory of random flights \cite{hugh1}. The scalar
potential, of course, will remain close to zero; it is only the vector
potential that accumulates these additions. Consequently, from now on we
will write $\langle A^{\mu} A_{\mu} \rangle_{\mathrm{th}} =
\langle {\mathbf A}^2 \rangle_{\mathrm{th}}$, and similarly for the
non-thermal part.

We are concerned with the ratio
\begin{equation}
p = \langle {\mathbf A}^2 \rangle_{\mathrm{nt}}/
\langle {\mathbf A}^2 \rangle_{\mathrm{th}} . \label{eq:p_ratio}
\end{equation}
Both the numerator and the denominator will decrease as the temperature
falls, but the numerator falls more slowly, so the ratio will build up
from zero until it becomes of order unity. At this point a Coleman-Weinberg
transition \cite{cole1} can take place and normal masses will appear.
These will have magnitude
$m^2 \approx \langle {\mathbf A}^2 \rangle_{\mathrm{nt}}$. A schematic
diagram illustrating this process can be found in Narlikar and Padmanabhan
\cite{nar2}, figure 10.1. When $\langle {\mathbf A}^2 \rangle_{\mathrm{nt}}$
exceeds $\langle {\mathbf A}^2 \rangle_{\mathrm{th}}$ a second minimum,
lower than the one at $m=0$, will develop in the effective potential.

\begin{figure}[h]
\centering
\ifthenelse {\lengthtest{\baselineskip > 16pt}}	
{
	\includegraphics[scale=0.9]{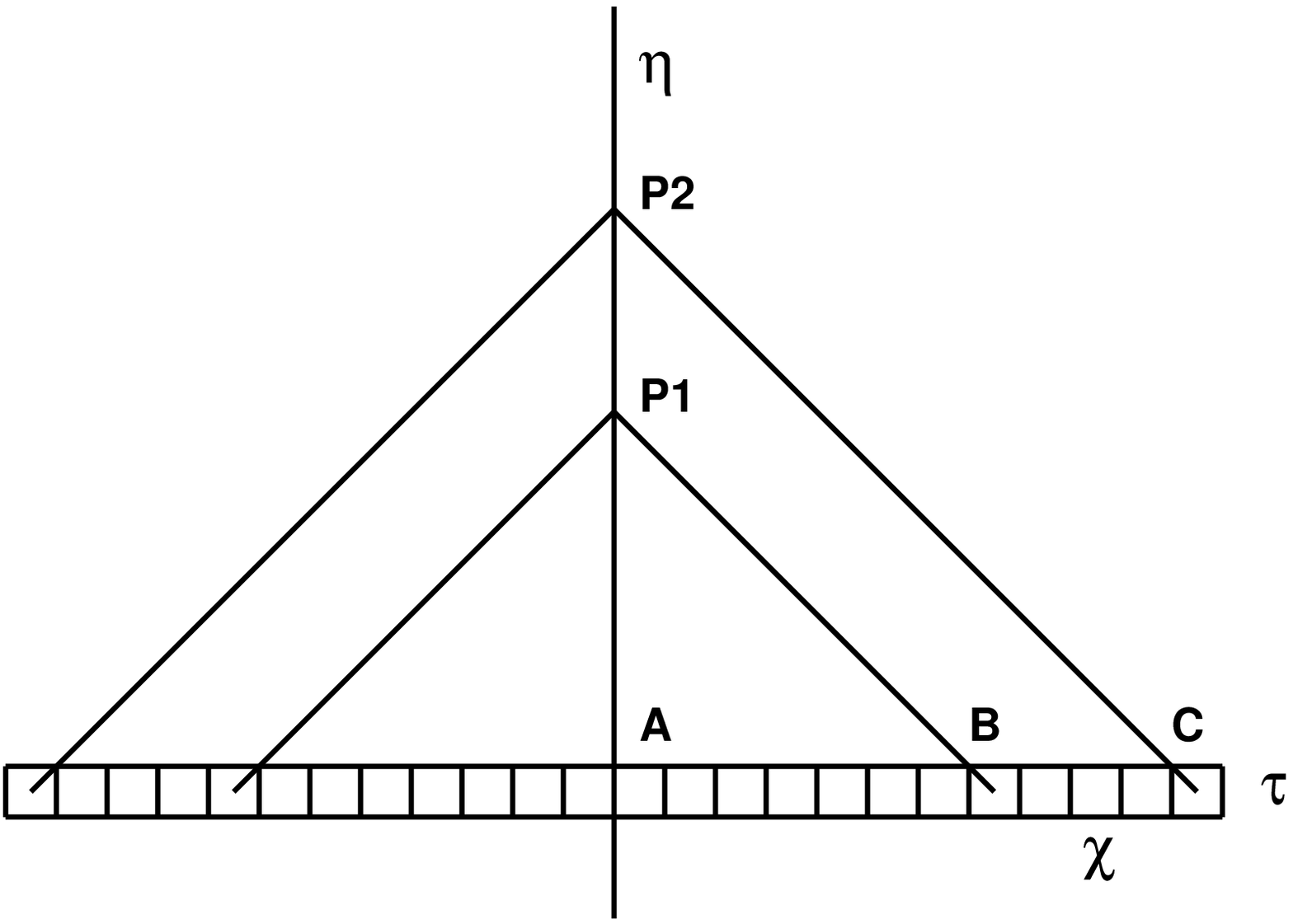}
} {	
	\includegraphics[width=3.375in]{msepfig5.eps}
}	
\caption{\label{fig:build} Buildup of
$\langle {\mathbf A}^2 \rangle_{\mathrm{nt}}$. Each box represents a
spherical shell of radius $\chi$, thickness $\mathrm{d} \chi$, duration
$\tau$. For clarity, only one time slice is shown; the plasma actually
fills the whole volume from $\eta = 0$ to the observation time.  }
\end{figure}

It is convenient at this point to change coordinates so the observation
point is at $\eta = \eta_p$, $\chi = 0$, and a general source point is at
$\eta$, $\chi$. Our formulae involve differences in $\eta$ and $\chi$, which
are unchanged by this shift of origin. We will consider $\eta$ and $t$ as
starting from zero at the time of minimum radius, $R_{\mathrm{min}}$, and
maximum temperature, $T_{\mathrm{max}}$. We can picture the buildup of
$\langle {\mathbf A}^2 \rangle_{\mathrm{nt}}$
with the help of figure \ref{fig:build}, where
the coordinates are $\eta$ (upwards) and $\chi$. The row of boxes at the
bottom of the diagram is at $\eta = 0$; this represents the surface of the
bubble. Each box has thickness $\mathrm{d} \chi$ and duration $\tau$. Two
observation points are shown, ${\mathrm{P}}_1$ and ${\mathrm{P}}_2$, at
times $\eta_1$ and $\eta_2$. The lines ${\mathrm{P}}_1$--B and
${\mathrm{P}}_2$--C represent the past light cones. All plasma particles
within these past light cones contribute to
$\langle {\mathbf A}^2 \rangle_{\mathrm{nt}}$. The contributions add
incoherently over times longer than the collision time, $\tau$, so we can
imagine the whole diagram divided into time slices of duration $\tau$, just
like the one shown for $\eta = 0$. A box at source time
$\eta_{\mathrm{s}}$ represents a spherical shell of volume
$4\pi R_{\mathrm{s}}^3 \sinh^2 \chi \, \mathrm{d} \chi$.

The number density of particles in this shell, for a thermal distribution, is
\begin{equation}
\nu_{\mathrm{s}} = 0.24 T_{\mathrm{s}}^3 , \label{eq:numdensity}
\end{equation}
and the number of particles in the shell is then
\begin{eqnarray}
\mathrm{d}N_{\mathrm{s}} & = & 4 \pi (0.24 T_{\mathrm{s}}^3)
R_{\mathrm{s}}^3 \sinh^2 \chi \, \mathrm{d} \chi \nonumber \\
  & \equiv & 4 \pi (0.24) A_{\mathrm{M}}^{3/4} \sinh^2 \chi \,
\mathrm{d} \chi .  \label{eq:numshell}
\end{eqnarray}
Here we have introduced, following Mannheim \cite{mann6}, the quantity
$A_{\mathrm{M}} = R^4 T^4$, which (at our current level of calculation)
will be constant during the expansion, and therefore does not need another
subscript, `s' or `p'.

Multiplying (\ref{eq:numshell}) and (\ref{eq:A2fit}) we get the contribution
to $A_2^2$ from the particles in the shell. We convert this to
$g^{\mu\nu} A_{\mu} A_{\nu}$ by multiplying by $g^{22}$:
\ifthenelse {\lengthtest{\baselineskip > 16pt}}	
{
	\begin{equation}
	4\pi (0.24) A_{\mathrm{M}}^{3/4} \sinh^2 \chi
	\, \mathrm{d} \chi \left(2q \tau \right)^2
	\langle V^2 \rangle \langle \sin^2 \theta \rangle 
	\frac{\tanh^2 (a \chi)}{R_{\mathrm{p}}^2 \sinh^2 \chi} .
	\label{exp:A2shell}
	\end{equation}
} {	
	\begin{eqnarray}
	4\pi (0.24) A_{\mathrm{M}}^{3/4} \sinh^2 \chi
	\, \mathrm{d} \chi \left(2q \tau \right)^2 \nonumber \\
	\times \langle V^2 \rangle \langle \sin^2 \theta \rangle 
	\frac{\tanh^2 (a \chi)}{R_{\mathrm{p}}^2 \sinh^2 \chi} .
	\label{exp:A2shell}
	\end{eqnarray}
}	

The average of $\sin^2 \theta$ over a sphere gives $2/3$. For the average of
$V^2$ we reason as follows: if the effective $m^2$ of the particles were
purely thermal, we could write $V^2 = 1/2$. But the effective mass
actually increases with time as the ratio $p$ from (\ref{eq:p_ratio})
increases from zero towards one. So a better estimate of $V^2$ is
$1/(2+p)$.

$\eta = 0$ at the surface of the bubble, and $\eta = \eta_{\mathrm{p}}$ at
the observation point. For some time slice at time $\eta$ between these two
limits,  we get the contribution of all spherical shells
to $\langle {\mathbf A}^2 \rangle_{\mathrm{nt}}$ by integrating from
$\chi = 0$ to the light cone at $\chi = \eta_{\mathrm{p}} - \eta$:
\begin{eqnarray}
\frac{8\pi (0.24)}{3} A_{\mathrm{M}}^{3/4} \left(\frac{1}{2+p}\right)
\left(2q \tau \right)^2 \int_0^{\eta_{\mathrm{p}} - \eta} \!\!
\mathrm{d} \chi \frac{\tanh^2 (a \chi)}{R_{\mathrm{p}}^2 } . \quad
\label{exp:A2slice}
\end{eqnarray}

We now have to sum over all time slices of duration $\tau$. The sum can be
converted to an integral by multiplying by $\mathrm{d} \eta/\tau$:
\ifthenelse {\lengthtest{\baselineskip > 16pt}}	
{
	\begin{equation}
	\frac{32\pi (0.24)}{3} A_{\mathrm{M}}^{3/4} q^2
	\int_0^{\eta_{\mathrm{p}}} \!\! \mathrm{d} \eta
	\left(\frac{1}{2+p}\right) \tau \int_0^{\eta_{\mathrm{p}} - \eta}
	\!\! \mathrm{d} \chi \frac{\tanh^2 (a \chi)}{R_{\mathrm{p}}^2 } .
	\label{exp:A2both}
	\end{equation}
} {	
	\begin{eqnarray}
	\frac{32\pi (0.24)}{3} A_{\mathrm{M}}^{3/4} q^2
	\int_0^{\eta_{\mathrm{p}}} \!\! \mathrm{d} \eta
	\left(\frac{1}{2+p}\right) \tau \nonumber \\
	\times \int_0^{\eta_{\mathrm{p}} - \eta}
	\!\! \mathrm{d} \chi \frac{\tanh^2 (a \chi)}{R_{\mathrm{p}}^2 } .
	\label{exp:A2both}
	\end{eqnarray}
}	

In section \ref{sec:screen} we defined $\tau = \tau_t / R_{\mathrm{s}} (t)$.
For $\tau_t$ we will simply use equation (3.24) from \cite{mont2}. The
factor $U^3$ in the numerator will be of order $(2+p)^{-3/2}$; the quantity
in the bracket in the denominator is of order unity and will be omitted. We
then have, in our notation,
$\tau_t = m^2(2+p)^{-3/2}/(8\pi \nu_{\mathrm{s}} q^4 \ln \Lambda)$.
For $m^2$ we will use the effective mass at time $t$. The purely thermal
value is $q^2 T_{\mathrm{s}}^2$, but we should multiply this by $(1+p)$ to
include the non-thermal part also. We have
$\nu_{\mathrm{s}} = 0.24 T_{\mathrm{s}}^3$, so
\begin{eqnarray}
\tau & = & \frac{1+p}
{8 \pi \ln \Lambda (0.24) T_{\mathrm{s}} q^2 R_{\mathrm{s}} (2+p)^{3/2} }
\nonumber \\
  & = & \frac{1+p}
{8 \pi \ln \Lambda (0.24) q^2 A_{\mathrm{M}}^{1/4} (2+p)^{3/2} } .
\label{eq:tau_est}
\end{eqnarray}

We can now get an expression for $p$ by using (\ref{eq:tau_est}) in
(\ref{exp:A2both}) and dividing by $T_{\mathrm{p}}^2$:
\ifthenelse {\lengthtest{\baselineskip > 16pt}}	
{
	\begin{eqnarray}
	p & = & \frac{4}{3 \ln \Lambda}
	A_{\mathrm{M}}^{3/4} q^2 \int_0^{\eta_{\mathrm{p}}}
	\!\! \mathrm{d} \eta
	\left(\frac{1+p}{ q^2 A_{\mathrm{M}}^{1/4} (2+p)^{5/2} } \right) 
	\int_0^{\eta_{\mathrm{p}} - \eta} \!\! \mathrm{d} \chi
	\frac{\tanh^2 (a \chi)}{R_{\mathrm{p}}^2 T_{\mathrm{p}}^2} \nonumber \\
	  & = & \frac{4}{3 \ln \Lambda}
	\int_0^{\eta_{\mathrm{p}}} \!\! \mathrm{d} \eta 
	\frac{1+p}{(2+p)^{5/2}}
	\int_0^{\eta_{\mathrm{p}} - \eta} \!\! \mathrm{d} \chi \tanh^2 (a \chi) 
	\nonumber \\
	  & = & \frac{4}{3 \ln \Lambda} \int_0^{\eta_{\mathrm{p}}} \!\!
	\mathrm{d} \eta \frac{1+p}{(2+p)^{5/2}}
	\left\{\eta_p - \eta - \tanh [a(\eta_p - \eta)]/a \right\} .
	\label{eq:p_exp_1}
	\end{eqnarray}
} {	
	\begin{eqnarray}
	p & = & \frac{4}{3 \ln \Lambda}
	A_{\mathrm{M}}^{3/4} q^2 \int_0^{\eta_{\mathrm{p}}}
	\!\! \mathrm{d} \eta
	\left(\frac{1+p}{ q^2 A_{\mathrm{M}}^{1/4} (2+p)^{5/2} } \right) 
	\nonumber \\
	  & & \times \int_0^{\eta_{\mathrm{p}} - \eta} \!\! \mathrm{d} \chi
	\frac{\tanh^2 (a \chi)}{R_{\mathrm{p}}^2 T_{\mathrm{p}}^2} \nonumber \\
	  & = & \frac{4}{3 \ln \Lambda}
	\int_0^{\eta_{\mathrm{p}}} \!\! \mathrm{d} \eta 
	\frac{1+p}{(2+p)^{5/2}}
	\int_0^{\eta_{\mathrm{p}} - \eta} \!\! \mathrm{d} \chi \tanh^2 (a \chi) 
	\nonumber \\
	  & = & \frac{4}{3 \ln \Lambda} \int_0^{\eta_{\mathrm{p}}} \!\!
	\mathrm{d} \eta \frac{1+p}{(2+p)^{5/2}} \nonumber \\
	  & & \times \left\{\eta_p - \eta - \tanh [a(\eta_p - \eta)]/a \right\} .
	\label{eq:p_exp_1}
	\end{eqnarray}
}	

This integral equation is surely not correct in detail, but it does provide
a general picture of the buildup of $p$ from $p=0$ at $\eta_p=0$.
We seek $\eta_{\mathrm{cw}}$, the value of $\eta_p$ for which $p=1$. Use
$a = 0.735$, as before, and for definiteness set $\ln \Lambda = 10$;
numerical solution then gives $\eta_{\mathrm{cw}} \approx 11$.

A remarkable feature of (\ref{eq:p_exp_1}) is that not only has
$A_{\mathrm{M}}$ disappeared, but also the coupling constant, $q$. The
long-range character of the Coulomb interaction, however, is still apparent
in $\ln \Lambda$. An equation of this sort might reasonably have been
expected to yield a value of $\eta_{\mathrm{cw}}$ that was either extremely
small, so the CW transition takes place almost immediately after the
formation of the bubble, or extremely large, so the transition never takes
place at all.  Instead we get a value of $\eta_{\mathrm{cw}}$ that is within
one or two orders of magnitude of unity.

As an example of the application of this result, we estimate, in appendix
\ref{app:Mannheim}, parameters in the model developed by Mannheim
\cite{mann6} for conformal gravity.

One might think that in setting up potentials by the ``classical
prescription'' of subsection \ref{subsec:gaugeinvar} we would find that at
any point a large value of $\langle \mathbf{A} \rangle$ would develop, as
well as $\langle {\mathbf A}^2 \rangle_{\mathrm{nt}}$. But this is not so,
once we treat the Universe as a quantum
mechanical system. We must avoid being too specific about the state of
distant matter, which we cannot investigate directly. We should not assume 
it is in a pure state; rather, it is a mixture, and all states compatible
with the conservation laws will be present in the density matrix. In many
situations, as emphasized by ter Haar \cite{terhaar}, it is a matter of
taste whether we adopt the statistical (ensemble) or quantum mechanical
interpretation of the density matrix. In the language of Bell \cite{bell1},
``or'' and ``and'' are equally acceptable. But this is not the case here;
the quantum mechanical interpretation is the appropriate one.
In an ordinary plasma in Minkowski space, only the  particles within the
Debye sphere will contribute to ${\mathbf A}$, and we do not have to concern
ourselves with mixtures. But propagation in a curved space obliges us to
consider very large numbers of particles at cosmological distances, and 
the introduction of a mixture is inevitable.

Like the distant Universe itself, $\langle {\mathbf A}^2 \rangle$ will be a
mixture, and the only meaningful quantities are averages. By symmetry,
$\langle {\mathbf A} \rangle$ will be zero, but
$\langle {\mathbf A}^2 \rangle$ will not, so
$q^2 \langle {\mathbf A}^2 \rangle$ can play the role of $m^2$.

\section{\label{sec:introcond}INTRODUCTION OF CONDUCTIVITY}

Up to this point we have been mainly concerned with potentials in the tail.
But we have now to look more closely at the electric and magnetic fields in
the main pulse for dipole propagation. These combine to give a Poynting
vector directed outwards. In flat space, this vector falls off like
$1/\chi^4$, which tends to zero even when integrated over the whole sphere.
When we go to curved space, however, the Poynting vector tends to
$(q V \sin \theta / \sinh \chi )^2$. The surface area of a sphere is 
$4 \pi R^2 \sinh^2 \chi$, so the integral of the Poynting vector tends to a
constant non-zero value. One effect of curvature is to continuously increase
the energy of the plasma.

It seems unlikely this extra energy will remain in the form of pulses like
those of figure \ref{fig:HPEPS}. We know such pulses propagate unchanged in
a flat space, but when they have traveled cosmological distances, so the
curvature becomes noticeable, we should expect a gradual thermalization.
We will model this thermalization by including a small, constant
conductivity, $\sigma$, in our equations. Note that $\sigma$ is not directly
related to the normal conductivity of the plasma, which (in our cosmological
units) would be very high. $\sigma$ is just a device to represent the slow
thermalization, and will have a value of order unity. For technical reasons,
which we explain below, we choose $\sigma = 1/(2\pi)$.

There are two reasons why we have to investigate the effect of $\sigma$:
\begin{enumerate}
	\item The thermalization of the pulses raises the temperature of the
	plasma, and so tends to reduce the value of the ratio $p$.
	\item The slow decline in the height of the pulses will probably reduce
	the value of $A$ in the tail, and this will also reduce the value of $p$.
\end{enumerate}

The gradual transfer of energy to the plasma implies that the product 
$R(\eta) T(\eta)$ will not remain constant, as in a simple expansion, but
will slowly increase. In this respect $\sigma$ simulates inflation, but 
on a much longer time scale.

\section{Inclusion of conductivity: fields}
\label{sec:incl_conduct}

The wave number, $k$, the permittivity, $\epsilon (n)$, and the
conductivity, $\sigma$, satisfy the dispersion relation
\begin{equation}
k^2 = n^2 \epsilon(n) = n(n + 4\pi i \sigma) . \label{eq:dispersion}
\end{equation}

By analogy with Maxwell's equations in flat space, the equations
(\ref{eq:dip1}), (\ref{eq:dip2}) and (\ref{eq:dip3}) for our
dipole fields become:
\begin{eqnarray}
i n \epsilon \sinh^2 \chi f_1 - 2 f_3 & = & 0 , \label{eq:dip1perm} \\
i n \epsilon f_2 - f_3^{\,\prime} & = & 0 , \label{eq:dip2perm} \\
f_1 - f_2^{\,\prime} + i n f_3 & = & 0 . \label{eq:dip3perm} 
\end{eqnarray}

\subsection{Magnetic field}
\label{subsec:conduct_H}

From these we obtain an equation for $f_3$ alone:
\begin{equation}
\frac{d^2 f_3}{d \chi^2} +
\left(n^2 \epsilon - \frac{2}{\sinh^2 \chi} \right) f_3 = 0 .
\label{eq:f3perm}
\end{equation}

Writing $k^2$ for $n^2 \epsilon$ this equation takes a familiar form; $f_3$
can be derived from the empty-space formula simply by writing $k$ for $n$
everywhere, except, of course, for the normalization function,
which we denote now by $C_3 (n,\sigma)$:
\begin{eqnarray}
f_3 & = &  \frac{C_3 (n,\sigma) \exp (ik\chi)}{2 u}
\left(1 - 2iku + u^2 \right) . \label{eq:f3out2perm}
\end{eqnarray}

$C_3 (n,\sigma)$ is most easily determined in flat space; this is sufficient
since we only have to consider small distances. The calculation is done in
Appendix \ref{app:normalize}; see (\ref{eq:normC3perm}).

\subsection{Electric fields}
\label{subsec:conduct_E}

(\ref{eq:dip1perm}) and (\ref{eq:f3out2perm}) give the radial component of
the electric field:
\ifthenelse {\lengthtest{\baselineskip > 16pt}}	
{
	\begin{eqnarray}
	f_1 & = & \frac{-i C_3 (n,\sigma ) \exp(ik \chi )
	(1-u^2)^2}{4(n+4 \pi i \sigma)u^3} \left(1 - 2iku + u^2 \right) .
	\label{eq:f1outperm}
	\end{eqnarray}
} {	
	\begin{eqnarray}
	f_1 & = & \frac{-i C_3 (n,\sigma ) \exp(ik \chi )
	(1-u^2)^2}{4(n+4 \pi i \sigma)u^3} \nonumber \\
	& & \times \left(1 - 2iku + u^2 \right) . \label{eq:f1outperm}
	\end{eqnarray}
}	

(\ref{eq:dip2perm}) and (\ref{eq:f3out2perm}) give the transverse component
of the electric field:
\ifthenelse {\lengthtest{\baselineskip > 16pt}}	
{
	\begin{eqnarray}
	f_2 & = & \frac{i C_3 (n,\sigma) \exp (ik \chi )}{4(n+4 \pi i \sigma) u^2}
	\left[(1-u^2)^2 - 2iku(1+u^2) - 4k^2 u^2 \right] . \label{eq:f2outperm}
	\end{eqnarray}
} {	
	\begin{eqnarray}
	f_2 & = &
	\frac{i C_3 (n,\sigma) \exp (ik \chi )}{4(n+4 \pi i \sigma) u^2}
	\nonumber \\
	& & \times \left[(1-u^2)^2 - 2iku(1+u^2) - 4k^2 u^2 \right] .
	\label{eq:f2outperm}
	\end{eqnarray}
}	

\subsection{Propagation of the magnetic field}
\label{subsec:propf12perm}

The Fourier transform of the magnetic field becomes
\ifthenelse {\lengthtest{\baselineskip > 16pt}}	
{
	\begin{equation}
	F_{12} = \frac{-3qV (n + 4\pi i \sigma ) N_1 (\theta)
	\sin (n\tau) \exp (ik\chi - in\eta)} {(3n + 8 \pi i \sigma ) n u}
	\left(1 - 2iku + u^2 \right) . \label{eq:F12perm}
	\end{equation}
} {	
	\begin{eqnarray}
	F_{12} & = & \frac{-3qV (n + 4\pi i \sigma ) N_1 (\theta)
	\sin (n\tau) \exp (ik\chi - in\eta)} {(3n + 8 \pi i \sigma ) n u}
	\nonumber \\
	& & \times \left(1 - 2iku + u^2 \right) . \label{eq:F12perm}
	\end{eqnarray}
}	

We can integrate around the pole at $n=0$ in the same way as before, except
that we have to respect the branch points of $k$, at $n=0$ and
$n=-4\pi i \sigma$. We have also to take account of the pole at
$n = -8 \pi i \sigma / 3$. A suitable contour is shown in figure
\ref{fig:CONTOUR2EPS}.

\ifthenelse {\lengthtest{\baselineskip > 16pt}}	
{
	\begin{figure}[h]
	\centering
	\includegraphics[scale=0.9]{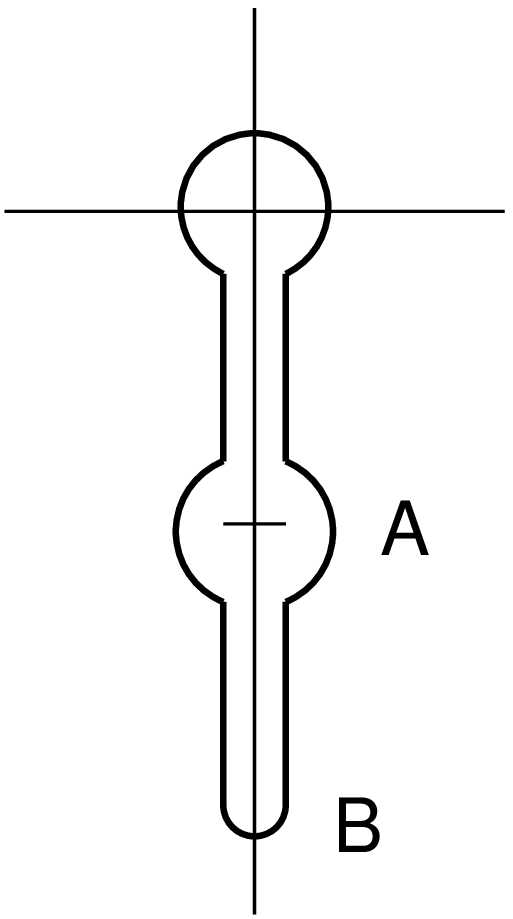}
	\caption{\label{fig:CONTOUR2EPS}Contour for computing the magnetic field,
	$F_{12}$, when conductivity is included. The points on the negative
	imaginary axis labeled $A$ and $B$ are at $n = -8\pi i \sigma / 3$ and
	$n = -4 \pi i \sigma$, respectively. The contour is traversed in a
	clockwise sense. }
\end{figure}
} {	
}	

The integrals are straightforward, and the resulting propagation of
$F_{12}$ is shown in figure \ref{fig:HP2EPS}. Notice that the pulses now
show a tail that represents a reflected wave. This is unlikely to be
significant in practice, because any irregularities in the plasma will tend
to disrupt the coherence of the wave as it converges on the origin.

We note here for future reference that when $\eta \gg 1$ we can derive a
simple asymptotic form for $F_{12}$ and $F_{20}$, using the fact that in
the integration around the contour the integrand is concentrated near $n=0$.
We just give the result:
\begin{eqnarray}
F_{12} & \approx & 3qV \sin \theta \sqrt{\frac{\sigma}{\eta + \tau }}
\exp \left( \frac{-\pi \sigma \chi^2 }{\eta + \tau } \right) ,
\label{eq:F12asymptotic} \\
F_{20} & \approx & F_{12} / (4 \pi \sigma ) .
\label{eq:F20asymptotic}
\end{eqnarray}

\ifthenelse {\lengthtest{\baselineskip > 16pt}}	
{
} {	
	\begin{figure}[t]
	\centering
	\includegraphics[height=2.5in]{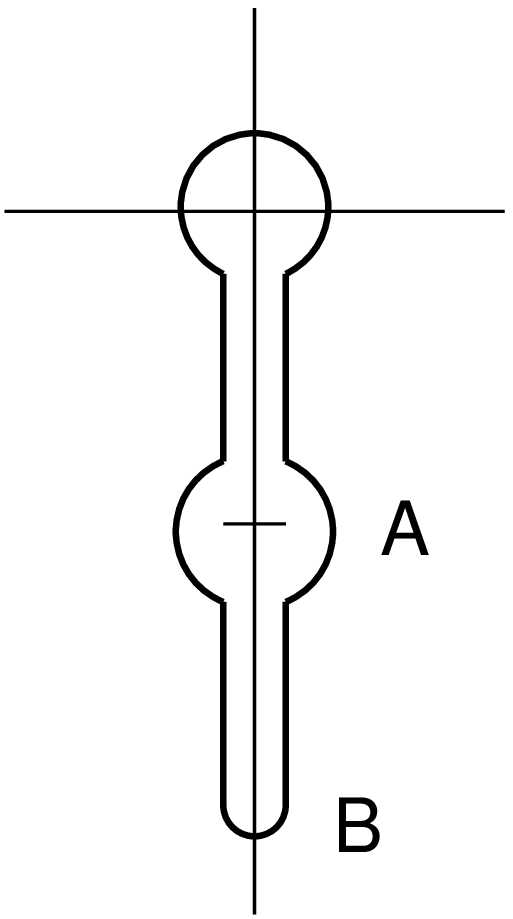}
	\caption{\label{fig:CONTOUR2EPS}Contour for computing the magnetic field,
	$F_{12}$, when conductivity is included. The points on the negative
	imaginary axis labeled $A$ and $B$ are at $n = -8\pi i \sigma / 3$ and
	$n = -4 \pi i \sigma$, respectively. The contour is traversed in a
	clockwise sense. }
	\end{figure}
}	

\ifthenelse {\lengthtest{\baselineskip > 16pt}}	
{
	\begin{figure}[ht]
	\centering
	\includegraphics[scale=0.9]{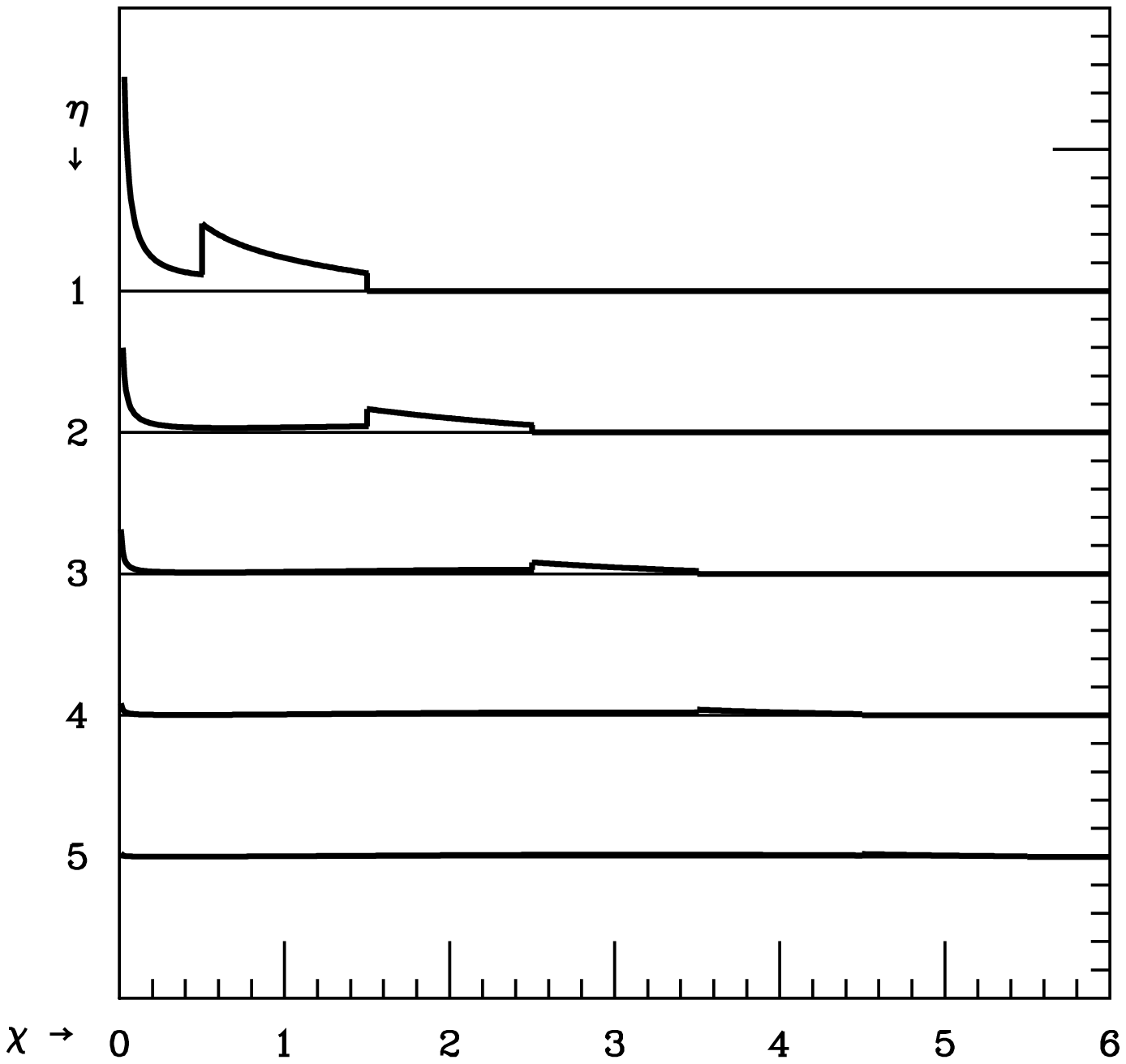}
} {	
	\begin{figure}[bb]
	\centering
	\includegraphics[width=3.375in]{msepfig7.eps}
}	
\caption{\label{fig:HP2EPS}Propagation of the magnetic field, $F_{12}$,
in the presence of conductivity, $\sigma = 1/2\pi$. }
\end{figure}

\section{Inclusion of conductivity: potentials}
\label{sec:conduct_A}

Here we follow the prescription of Landau and Lifshitz \cite{landau}:
use the same equations relating potentials and fields as in empty space,
i.e. (\ref{eq:heq1}), (\ref{eq:heq2}) and (\ref{eq:heq3}), but modify the
Lorenz condition by including the permittivity in the $h_0$ term:
\begin{equation}
i n \epsilon \sinh^2 \chi h_0 +
\frac{d}{d\chi} \left[\sinh^2 \chi h_1 \right] - 2 h_2 = 0 .
\label{eq:heq4perm}
\end{equation}

\subsection{Scalar potential}
\label{subsec:conduct_h0}

From these equations, as before, we can derive an equation for $h_0$ alone:
\begin{equation}
\frac{d^2 h_0}{d \chi^2} + 2 \coth \chi \frac{d h_0}{d \chi}
+ \left(n^2 \epsilon - \frac{2}{\sinh^2 \chi} \right) h_0 = 0 .
\label{eq:h0perm}
\end{equation}
This is the same equation as we obtained for empty space, except that, just
as for the magnetic field, in place of $n^2$ we must write
$k^2 \equiv n^2 \epsilon$. (\ref{eq:h0out1}) applies as before, provided we
write $\alpha = \sqrt{k^2 - 1}$, and use a normalization function $C_4
(n,\sigma )$ in place of $C_2 (n)$. 

This normalization function can be found from (\ref{eq:heq2}) and
(\ref{eq:f1outperm}), in the limit $\chi \rightarrow 0$. We get, in place
of (\ref{eq:normC2}),
\begin{equation}
C_4 (n,\sigma ) = \frac{iC_3 (n,\sigma )}{4(n + 4 \pi i \sigma )} .
\label{eq:normC4perm}
\end{equation}

For the Fourier synthesis we need, in general, to use a more complicated
contour than the one shown in figure \ref{fig:CONTOUR2EPS}, because we
have to take account of the branch points of both $k$ and $\alpha$. With our
choice of $\sigma = 1/2\pi$, however, the branch points of $\alpha$ coalesce
into a single point at $n=-i$, and the contour of figure
\ref{fig:CONTOUR2EPS} is adequate.

\subsection{Vector potential}
\label{subsec:conduct_h2}

(\ref{eq:heq2a}) holds, as in the case of zero conductivity.
The first term on the right side of (\ref{eq:heq2a}) comes from
(\ref{eq:f2outperm}):
\ifthenelse {\lengthtest{\baselineskip > 16pt}}	
{
	\begin{equation}
	\frac{-i f_2}{n} =
	\frac{C_3 (n,\sigma) \exp (ik \chi )}{4n(n+4 \pi i \sigma) u^2}
	\left[(1-u^2)^2 - 2iku (1+u^2) - 4k^2u^2 \right] .
	\end{equation}
} {	
	\begin{eqnarray}
	\frac{-i f_2}{n} & = &
	\frac{C_3 (n,\sigma) \exp (ik \chi )}{4n(n+4 \pi i \sigma) u^2}
	\nonumber \\
	& & \times \left[(1-u^2)^2 - 2iku (1+u^2) - 4k^2u^2 \right] . \quad
	\end{eqnarray}
}	

The second term on the right side of (\ref{eq:heq2a}) comes from
(\ref{eq:h0out1}):
\ifthenelse {\lengthtest{\baselineskip > 16pt}}	
{
	\begin{equation}
	\frac{i h_0}{n} = \frac{i C_4 (n,\sigma) \exp(i\alpha \chi )
	(1-u^2)}{n u^2} \left( 1 - 2i \alpha u + u^2 \right) .
	\end{equation}
} {	
	\begin{eqnarray}
	\frac{i h_0}{n}  & = & \frac{i C_4 (n,\sigma) \exp(i\alpha \chi )
	(1-u^2)}{n u^2} \nonumber \\
	& & \times \left( 1 - 2i \alpha u + u^2 \right) .
	\end{eqnarray}
}	

The two normalization functions are related by (\ref{eq:normC4perm}),
and we also know $k^2 = n(n+4 \pi i \sigma)$. Combining, we get:
\begin{eqnarray}
h_2 & = & \frac{C_3 (n,\sigma)}{4n(n+4 \pi i \sigma) u^2} 
\left( {\cal K_{\mathrm \sigma} } +{\cal A_{\mathrm \sigma} } \right) ,
\label{eq:heq2bperm} \\
{\cal K_{\mathrm \sigma}} & = & e^{i k \chi } 
\left[(1-u^2)^2 - 2iku (1+u^2) - 4k^2u^2 \right] , \quad 
\label{eq:calKperm} \\ 
{\cal A_{\mathrm \sigma}} & = & e^{i \alpha \chi } (1-u^2)
\left( -1 + 2i\alpha u - u^2 \right) . \label{eq:calAperm}
\end{eqnarray}

\subsection{Propagation of the vector potential}
\label{subsec:conduct_propA2}

Combining (\ref{eq:heq2bperm}) with (\ref{eq:Adef2}), (\ref{eq:normC3perm})
and (\ref{eq:dipstrength}), and setting $-8\pi i \sigma/3 = n_p$, we get
\begin{equation}
A_2 = \frac{-q V N_1 (\theta) \sin (n \tau) e^{-in \eta}}
{2 (n - n_p ) n^2 u^2}
\left( {\cal K_{\mathrm \sigma} } +{\cal A_{\mathrm \sigma} } \right) .
\label{eq:A2basicperm}
\end{equation}

We note that going from (\ref{eq:A2basic}) to (\ref{eq:A2basicperm})
takes three simple steps:
\begin{enumerate}
	\item Rename ${\cal N}$ and ${\cal A}$; call them ${\cal K}_{\sigma}$
	and ${\cal A}_{\sigma}$, respectively.
	\item In the final parenthesis, substitute $k$ for $n$ in both
	${\cal K}_{\sigma}$ and ${\cal A_{\sigma}}$; this includes redefining
	$\alpha$ to be $\sqrt{k^2 - 1}$ rather than $\sqrt{n^2 - 1}$.
	\item In the prefactor, change $n^3$ in the denominator to
	$n^2 (n - n_p)$.
\end{enumerate}

(\ref{eq:A2basicperm}) becomes (\ref{eq:A2basic}) in the limit
$\sigma \rightarrow 0$, as it should.

We transform to $\eta$, $\chi$, $\theta$, $\phi$ coordinates as before, by
dividing (\ref{eq:A2basicperm}) by $2\pi$ and integrating clockwise around
the contour of figure \ref{fig:CONTOUR2EPS}. The result is shown in figure
\ref{fig:AT2EPS}.
 
\begin{figure}[ht]
\centering
\ifthenelse {\lengthtest{\baselineskip > 16pt}}	
{
	\includegraphics[scale=1.0]{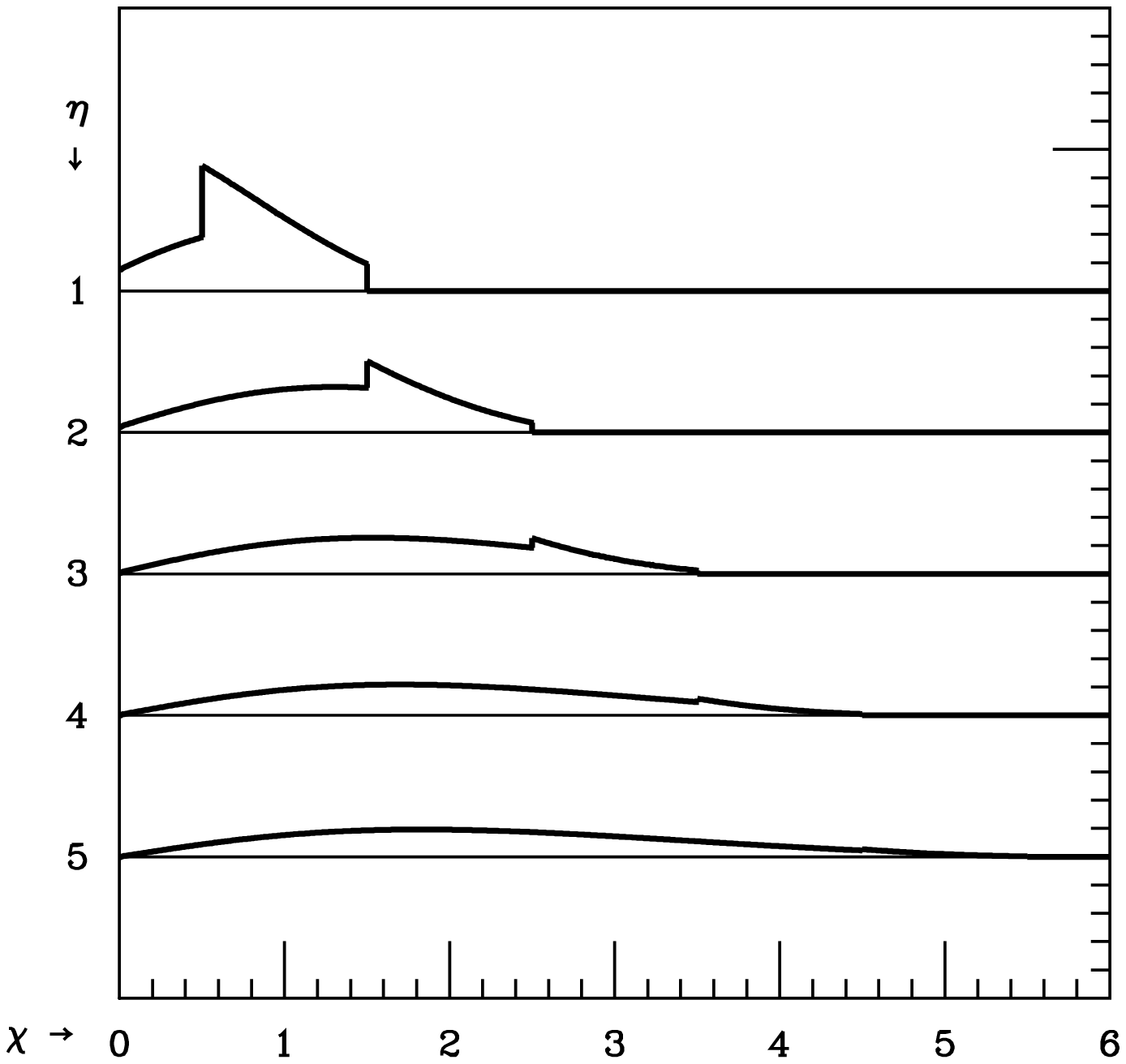}
} {	
	\includegraphics[width=3.0in]{msepfig8.eps}
}	
\caption{Propagation of the vector potential, $A_2$, in the presence of
conductivity, $\sigma = 1/2\pi$. }
\label{fig:AT2EPS}
\end{figure}

\subsection{Asymptotic form of the vector potential}
\label{subsec:A2asymptotic}

We can show easily that in the limit $\eta \rightarrow \infty$,
$A_2$ satisfies a diffusion equation. The main terms come from the integral
down the cut, and in this limit the integrand is concentrated close to
$n=0$. This makes it possible, as for the fields, to obtain a simple
expression for the asymptotic form:

\ifthenelse {\lengthtest{\baselineskip > 16pt}}	
{
	\begin{eqnarray}
	A_2 & \approx & \frac{3 q V \tau \sin \theta }{16 \pi}
	F( \sigma, \chi, \eta ) , \label{eq:A2asymptotic_2} \\
	F( \sigma, \chi, \eta ) & = & \left\{ \frac{(1-u^2)^2}{ \sigma u^2 }
	{\mathrm{erf}} \left(\chi \sqrt{\frac{\pi \sigma}{\eta}} \right)
	\right.  \nonumber \\
	  & & \left. {} - \frac{4 }{\sigma^{1/2} \eta^{3/2} u}
	\left[ (1+u^2)\eta + 4\pi \sigma u \chi \right] 
	\exp \left(\frac{-\pi \sigma \chi^2 }{\eta} \right) \right\} .
	\label{eq:Fasymptotic_2}
	\end{eqnarray}
} {	
	\begin{eqnarray}
	A_2 & \approx & \frac{3 q V \tau \sin \theta }{16 \pi}
	F( \sigma, \chi, \eta ) , \label{eq:A2asymptotic_2} \\
	F( \sigma, \chi, \eta ) & = & \left\{ \frac{(1-u^2)^2}{ \sigma u^2 }
	{\mathrm{erf}} \left(\chi \sqrt{\frac{\pi \sigma}{\eta}} \right)
	- \frac{4 }{\sigma^{1/2} \eta^{3/2} u} \right. \nonumber \\
	  & & \hspace{-4em} \left. {} \times
	\left[ (1+u^2)\eta + 4\pi \sigma u \chi \right] 
	\exp \left(\frac{-\pi \sigma \chi^2 }{\eta} \right) \right\} . \,
	\label{eq:Fasymptotic_2}
	\end{eqnarray}
}	

When $\chi$ is also large, of order $\sqrt{\eta}$, the term involving
the error function is negligible compared to the others.

In figure \ref{fig:AT3EPS} we plot $\sqrt{\eta} A_2$ against
$\xi = \chi/\sqrt{\eta}$, with $A_2$ computed using the asymptotic formula.
We see $\sqrt{\eta} A_2$ tending to a constant form; with our choice of
abscissa, this is a Gaussian for large $\chi$ and $\eta$. For $\eta \ge 4$,
this graph is indistinguishable from the one using integration around the
contour.

\begin{figure}[ht]
\centering
\ifthenelse {\lengthtest{\baselineskip > 16pt}}	
{
	\includegraphics[scale=1.0]{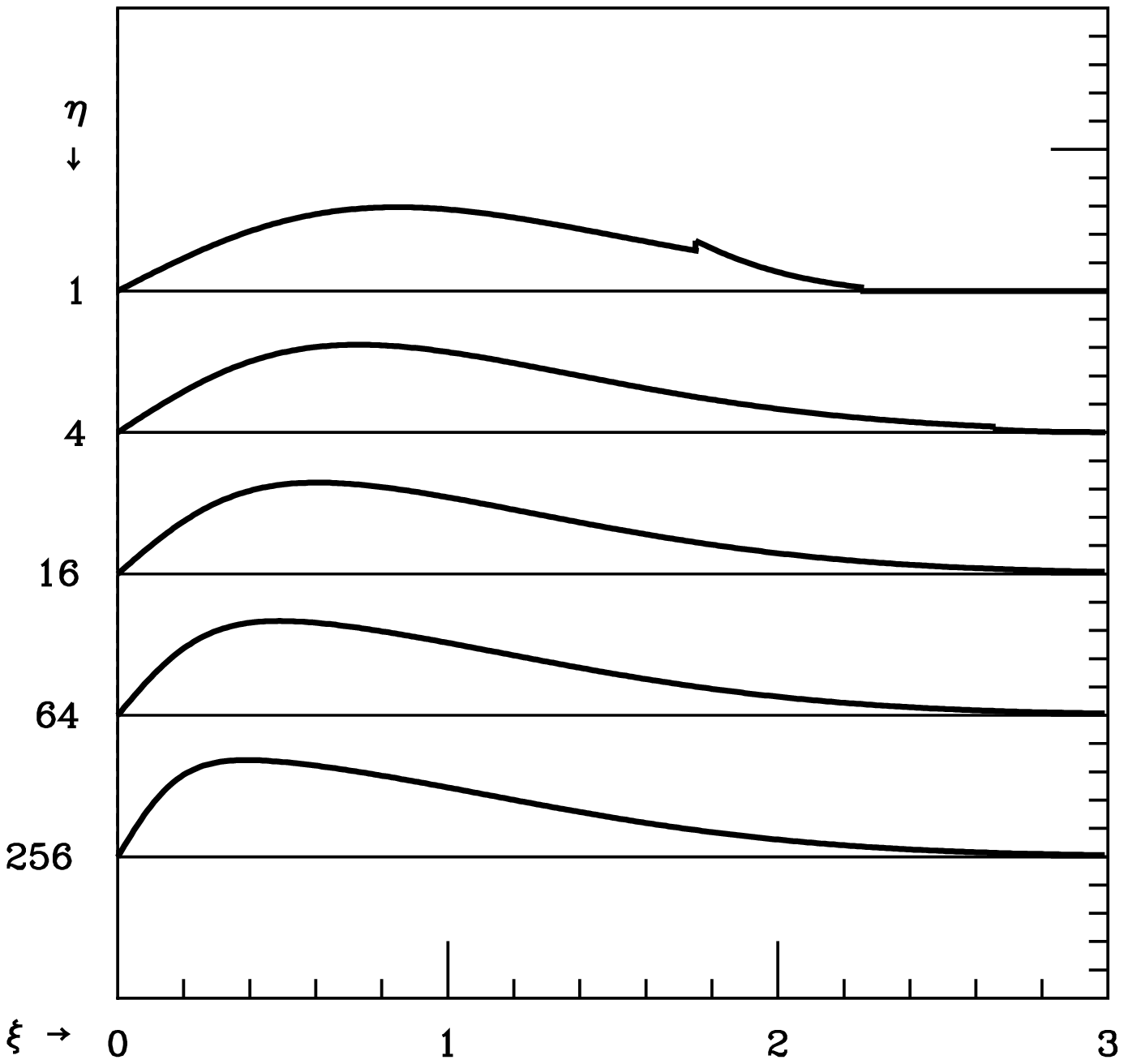}
} {	
	\includegraphics[width=3.0in]{msepfig9.eps}
}	
\caption{$\sqrt{\eta} A_2$ plotted against $\xi = \chi /\sqrt{\eta}$,
to demonstrate the asymptotic form. In this graph we use the asymptotic
form (\ref{eq:A2asymptotic_2}) for the tail. }
\label{fig:AT3EPS}
\end{figure}

\section{The energy-transfer equation}
\label{sec:energy_transfer}

We will now investigate the effect of a non-zero conductivity on the
temperature of the plasma. We start from Weinberg \cite{wein2},
equation (4.7.9):
\begin{equation}
T^{\mu\nu}_{\;\;\; ;\mu} = \frac{1}{\sqrt{g}}
\frac{\partial}{\partial x^{\mu}} \left( \sqrt{g} T^{\mu\nu} \right) +
\Gamma^{\nu}_{\mu\lambda} T^{\mu\lambda} . \label{eq:wein4_7_9}
\end{equation}
This expression is zero for any value of $\nu$, but the most important one
for us is $\nu = 0$. We treat the plasma as a perfect fluid at rest in
FRW coordinates $\eta$, $\chi$, $\theta$, $\phi$, so that the 4-velocity
$\left( U^0 , U^i \right)$ has $U^i = 0$, $i=1,2,3$. Since
$g_{\mu \nu } U^{\mu} U^{\nu} = -1$ and $g_{00} =  R^2 (\eta)$, 
$U^0 = 1/R$.

Multiply (\ref{eq:wein4_7_9}) by $\sqrt{g}\,\mathrm{d}^4 x\,U_{\nu}$,
sum over $\nu$, and set the resulting expression equal to zero; we get
\begin{equation}
\sqrt{g}\, \mathrm{d}^4 x\, U_0 T^{\mu 0}_{\;\;\; ;\mu} = 0 .
\label{eq:econ1}
\end{equation}
The left-hand side of this equation is a coordinate scalar with
$\sqrt{g} \mathrm{d}^4 x$ an element of proper volume. Expanding
(\ref{eq:econ1}):
\ifthenelse {\lengthtest{\baselineskip > 16pt}}	
{
	\begin{eqnarray}
	\sqrt{g}\, \mathrm{d}^4 x\, U_0 \left[ \frac{1}{\sqrt{g}}
	\frac{\partial}{\partial x^0 } \left( \sqrt{g}\, T^{00} \right) 
	+ \frac{1}{\sqrt{g}} \frac{\partial}{\partial x^i } 
	\left( \sqrt{g}\, T^{i 0} \right) + \Gamma^{0}_{ij} T^{ij}
	+ \Gamma^{0}_{00} T^{00} \right] = 0 . \label{eq:econ2}
	\end{eqnarray}
} {	
	\begin{eqnarray}
	{} & & \sqrt{g}\, \mathrm{d}^4 x\, U_0 \left[ \frac{1}{\sqrt{g}}
	\frac{\partial}{\partial x^0 } \left( \sqrt{g}\, T^{00} \right) 
	+ \frac{1}{\sqrt{g}} \frac{\partial}{\partial x^i } 
	\left( \sqrt{g}\, T^{i 0} \right) \right. \nonumber \\
	{} & & \left. {} + \Gamma^{0}_{ij} T^{ij}
	+ \Gamma^{0}_{00} T^{00} \right] = 0 . \label{eq:econ2}
	\end{eqnarray}
}	

From this we can derive an integrodifferential equation for  $T^{00}$.
Details are given in appendix \ref{app:energy_transfer}. The resulting
equation is most simply written in terms of
$A_{\mathrm{M}} (\eta) = R^4 (\eta) T^4 (\eta) \,\mathrm{(eV)^4 \cdot m^4}$.
A similar quantity was introduced in section \ref{sec:mass_1}, but there it
could be treated simply as a constant. Here we calculate its variation with
time:
\ifthenelse {\lengthtest{\baselineskip > 16pt}}	
{
	\begin{eqnarray}
	{} & & \dot{A}_M (\eta ) -
	P_4 \int_0^{\eta} \mathrm{d} \eta^{\,\prime}
	A_{M}^{3/4} (\eta^{\,\prime})
	\frac{\sigma (\eta - \eta^{\,\prime})}{2 + p(\eta )}
	\exp \left[ -2 \pi \sigma (\eta - \eta^{\,\prime}) \right] = 0 ,
	\label{eq:econ6} \\
	{} & & P_4 = 3.97 \times 10^{-7}\,\mathrm{eV \cdot m} .
	\label{eq:P4def} 
	\end{eqnarray}
} {	
	\begin{eqnarray}
	{} & & \dot{A}_M (\eta ) -
	P_4 \int_0^{\eta} \mathrm{d} \eta^{\,\prime}
	A_{M}^{3/4} (\eta^{\,\prime})
	\frac{\sigma (\eta - \eta^{\,\prime})}{2 + p(\eta )} \nonumber \\
	 & & \times \exp \left[ -2 \pi \sigma (\eta - \eta^{\,\prime}) \right]
	= 0 , \label{eq:econ6} \\
	{} & & P_4 = 3.97 \times 10^{-7}\,\mathrm{eV \cdot m} .
	\label{eq:P4def} 
	\end{eqnarray}
}	

Note that because of the appearance of the function $p(\eta )$, this equation
by itself is not sufficient to calculate $A_{\mathrm{M}}$. We obtain an
independent equation connecting these two functions in the next section.

\section{\label{sec:mass_2}MASS GENERATION: SECOND CALCULATION}
 
We will now obtain an equation for the ratio $p (\eta)$ analogous to
(\ref{eq:p_exp_1}), but based on the asymptotic form
(\ref{eq:A2asymptotic_2}) for the tail of the vector potential. In our
previous calculation, where the tail of $A_2$ reached a constant asymptotic
value, we could be sure that $p$ would rise monotonically; the only question
was whether there would be factors of $A_{\mathrm{M}}$ that would make the
increase far too fast or far too slow. (This is just what happens, for
example, if we use quadrupole rather than dipole potentials.) In this second
calculation, we can anticipate that there will not be any unwanted factors
of $A_{\mathrm{M}}$, but only a detailed calculation will show whether $p$
continues to rise with $\eta$.

The analog of (\ref{eq:numshell}) is
\begin{equation}
\mathrm{d}N_{\mathrm{s}} = 4 \pi (0.24)
A_{\mathrm{M}}^{3/4} (\eta_{\mathrm{s}} )
\sinh^2 \chi \, \mathrm{d} \chi .  \label{eq:numshell_2}
\end{equation}

To get the contribution to $A_2^2$ at $\eta_{\mathrm{p}}$ from the
particles in the shell, we multiply (\ref{eq:numshell_2}) by
$A_2^2 [ \sigma, \chi, (\eta_{\mathrm{p}} - \eta_{\mathrm{s}} )]$,
with $A_2 ( \sigma, \chi, \eta )$ given by (\ref{eq:A2asymptotic_2}).

Proceeding as before we get the analog of (\ref{exp:A2both}):
\ifthenelse {\lengthtest{\baselineskip > 16pt}}	
{
	\begin{eqnarray}
	\frac{2\pi (0.24) q^2 }{3 R^2 (\eta_{\mathrm{p}}) }
	\int_0^{\eta_{\mathrm{p}}} \mathrm{d} \eta_{\mathrm{s}}
	\left(\frac{ A_{\mathrm{M}}^{3/4} (\eta_{\mathrm{s}}) }
	{2+p (\eta_{\mathrm{s}})}\right) \tau 
	\int_0^{\eta_{\mathrm{p}} - \eta_{\mathrm{s}}} \!\! \mathrm{d} \chi
	F^2 \left[\sigma,\chi,(\eta_{\mathrm{p}} - \eta_{\mathrm{s}})\right] , \quad
	\label{exp:A2both_2}
	\end{eqnarray}
} {	
	\begin{eqnarray}
	{} & & \frac{2\pi (0.24) q^2 }{3 R^2 (\eta_{\mathrm{p}}) }
	\int_0^{\eta_{\mathrm{p}}} \mathrm{d} \eta_{\mathrm{s}}
	\left(\frac{ A_{\mathrm{M}}^{3/4} (\eta_{\mathrm{s}}) }
	{2+p (\eta_{\mathrm{s}})}\right) \tau \nonumber \\
	{} & & \times 
	\int_0^{\eta_{\mathrm{p}} - \eta_{\mathrm{s}}} \!\! \mathrm{d} \chi
	F^2 \left[\sigma,\chi,(\eta_{\mathrm{p}} - \eta_{\mathrm{s}})\right] ,
	\label{exp:A2both_2}
	\end{eqnarray}
}	
with $F_0$ given by (\ref{eq:Fasymptotic_2}).

The collision time, $\tau$, will be given by
\begin{equation}
\tau = \frac{(1+p)}{8 \pi \ln \Lambda (0.24) q^2
A_{\mathrm{M}}^{1/4} (\eta_{\mathrm{s}}) (2+p)^{3/2} } .
\label{eq:tau_est_2}
\end{equation}

We can now get an expression for $p$ by using (\ref{eq:tau_est_2}) in
(\ref{exp:A2both_2}) and dividing by $T^2 (\eta_{\mathrm{p}})$:
\ifthenelse {\lengthtest{\baselineskip > 16pt}}	
{
	\begin{eqnarray}
	p (\eta_{\mathrm{p}}) & = &
	\frac{3 }{256 \pi^2 \ln \Lambda A_{\mathrm{M}}^{1/2}(\eta_{\mathrm{p}}) }
	\int_0^{\eta_{\mathrm{p}}} \!\! \mathrm{d} \eta_{\mathrm{s}}
	\left( \frac{ \left[1+p (\eta_{\mathrm{s}})\right]
	A_{\mathrm{M}}^{1/2} (\eta_{\mathrm{s}}) }
	{\left[2+p (\eta_{\mathrm{s}})\right]^{5/2} }\right) 
	\int_0^{\eta_{\mathrm{p}} - \eta_{\mathrm{s}}} \!\! \mathrm{d} \chi
	F^2 \left[\sigma,\chi,(\eta_{\mathrm{p}} - \eta_{\mathrm{s}})\right] ,
	\qquad \label{eq:pinteqn_2}
	\end{eqnarray}
} {	
	\begin{eqnarray}
	p (\eta_{\mathrm{p}}) & = &
	\frac{3 }{256 \pi^2 \ln \Lambda A_{\mathrm{M}}^{1/2}(\eta_{\mathrm{p}}) }
	\nonumber \\
	  & & \times \int_0^{\eta_{\mathrm{p}}} \!\! \mathrm{d} \eta_{\mathrm{s}}
	\left( \frac{ \left[1+p (\eta_{\mathrm{s}})\right]
	A_{\mathrm{M}}^{1/2} (\eta_{\mathrm{s}}) }
	{\left[2+p (\eta_{\mathrm{s}})\right]^{5/2} }\right) 
	\nonumber \\
	  & & \times \int_0^{\eta_{\mathrm{p}} - \eta_{\mathrm{s}}}
	\!\! \mathrm{d} \chi F^2 \left[\sigma,\chi,(\eta_{\mathrm{p}}
	- \eta_{\mathrm{s}})\right] ,  \label{eq:pinteqn_2}
	\end{eqnarray}
}	
with $F_0$ given by (\ref{eq:Fasymptotic_2}).

\subsection{\label{subsec:pandAM}Calculation of $\bm{p}$ and
$\bm{A_{\mathrm{M}}}$}

As before, we set $\ln \Lambda = 10$ for definiteness. Simultaneous
numerical solution of (\ref{eq:econ6}) and (\ref{eq:pinteqn_2}) is
straightforward, but we need to choose a starting value for
$A_{\mathrm{M}}$. We have used two extreme values:
\begin{enumerate}
	\item $A_{\mathrm{M}}$ starts at
	$2.5 \times 10^{-26}\, \mathrm{eV^4 \cdot m^4}$. This is a ``natural''
	value, in the sense that the Universe begins with the typical wavelength
	of the radiation approximately equal to the radius, $R$. $p$ becomes 
	equal to unity around $\eta_{\mathrm{cw}} = 1650$.
	\item $A_{\mathrm{M}}$ starts at 
	$2.5 \times 10^{74}\, \mathrm{eV^4 \cdot m^4}$. This is approximately
	equal to the observed value today. $p$ becomes equal to unity around
	$\eta_{\mathrm{cw}} = 480$.
\end{enumerate}

These values of $\eta_{\mathrm{cw}}$ are significantly larger than the
previous value of $11$ when $\sigma = 0$, but they are still within a few
orders of magnitude of unity. The most important feature of this calculation
is that it shows $p(\eta)$ does continue to rise, even when $\sigma \ne 0$,
so a CW transition will still eventually become possible.

\section{\label{sec:conclude}CONCLUSION}

We have considered the problem of the generation of a particle mass scale in
a theory in which this scale is not determined by the gravitational
constant, through $m_{\mathrm{planck}}$. Conformal gravity \cite{mann6}
may turn out to be a theory of this type. We find that in such a theory
distant matter is significant in an unexpected way, and \textit{the mass
scale is to a large extent classically determined}. Since the mass scale
develops through the agency of the classical vector potential, it is bound
to appear in any model universe containing charged particles, providing the
underlying space has sufficient curvature and allows enough time.

Many questions remain, among them the following:
\begin{enumerate}
	\item Will a CW transition necessarily take place as the ratio
	$\langle {\mathbf A}^2 \rangle_{\mathrm{nt}}/
	\langle {\mathbf A}^2 \rangle_{\mathrm{th}}$ approaches unity?
	\item Is the tail of ${\mathbf A}$ important only in the early Universe,
	or are there other places, regions of high gravitational fields, where
	its effects can be observed even now?
	\item How does the real complicated early plasma determine such things
	as the collision time?
	\item Finally, perhaps most important, can we really expect
	${\mathbf A}$ to propagate over cosmological distances? We have taken
	findings from ordinary plasma theory and used them in the very
	different circumstances of the early Universe.
\end{enumerate}

These considerations are beyond the scope of this paper, which is solely
concerned with the interplay, in a simple model, of quantum mechanics
(density matrix, the Lagrangian for a scalar field, the Coleman-Weinberg
transition) and the classical equations of propagation of the ordinary
vector potential in a FRW space of negative curvature.

\begin{acknowledgments}
We acknowledge helpful correspondence with Bryce DeWitt, Leonard Parker,
Stephen Fulling, Don Melrose, David Montgomery and Philip Mannheim. We
also wish to thank the chairman and faculty of the department of physics
at Washington University for providing an office and computer support for
a retired colleague. Cosmic space may be infinite, but office space is at
a premium.
\end{acknowledgments}

\ifthenelse {\lengthtest{\baselineskip > 16pt}}	
{
	\pagebreak
} {	
}	

\appendix
\section{\label{app:H-G}HYPERGEOMETRIC FUNCTIONS}

Hypergeometric functions are usually difficult to handle because they depend
on three parameters. For all the hypergeometric functions we need in this 
paper, however, the parameters $a$, $b$, and $c$ can be expressed in terms 
of two numbers, an integer, $d$, and a quantity $\alpha$ which is complex
in general:
\begin{equation}
a = d + i\alpha;\,\,\, b = d - i\alpha;\,\,\, c = d + 1/2 .
\end{equation}

Hypergeometric functions that satisfy this condition can be expressed in 
terms of elementary functions. For proof we use formulae from
Abramowitz and Stegun (\cite{absteg}, hereafter AS). Define
$\mu = (1 - \cosh(\chi))/2$. Then, for $d=0$, AS (15.1.17) gives, in our
notation,
$F(i\alpha,-i\alpha; 1/2; \mu) = F[i\alpha,-i\alpha; 1/2; \sin^2(i\chi/2)]
= \cos (\alpha \chi )$. Other functions can be obtained by successive
differentiation or integration using AS (15.2.1). If $\alpha$ happens to be
purely real or purely imaginary the hypergeometric functions are real.

For dipole propagation the only functions we need are those with $d=-1$
and $d=2$.  They are listed below, together with the functions with
intermediate values of $d$. We can replace the argument $\mu$ by $\chi$:
\ifthenelse {\lengthtest{\baselineskip > 16pt}}	
{
	\begin{eqnarray}
	F(-1 + i\alpha, -1 -i\alpha; -1/2; \chi) & = &
	\cosh \chi \cos(\alpha\chi)
	+ \alpha\sinh \chi \sin(\alpha\chi) , \label{eq:hgdip1} \\
	F(i\alpha, -i\alpha; 1/2; \chi) & = & \cos(\alpha\chi) , \\
	F(1 + i\alpha, 1 -i\alpha; 3/2; \chi) & = &  
	\frac{\sin(\alpha\chi)}{\alpha\sinh \chi } , \\
	F(2 + i\alpha, 2 -i\alpha; 5/2; \chi) & = &
	\frac{-3[\alpha\sinh \chi \cos(\alpha\chi) 
	- \cosh \chi \sin(\alpha\chi)]} {\alpha(1 + \alpha^2)\sinh^3 \chi } .
	\label{eq:hgdip2} 
	\end{eqnarray}
} {	
	\begin{eqnarray}
	F(-1 + i\alpha, -1 -i\alpha; -1/2; \chi) & = & \nonumber \\
	  & & \hspace{-14em} \cosh \chi \cos(\alpha\chi)
	+ \alpha\sinh \chi \sin(\alpha\chi) , \label{eq:hgdip1} \\
	F(i\alpha, -i\alpha; 1/2; \chi) & = & \cos(\alpha\chi) , \\
	F(1 + i\alpha, 1 -i\alpha; 3/2; \chi) & = &  
	\frac{\sin(\alpha\chi)}{\alpha\sinh \chi } , \\
	F(2 + i\alpha, 2 -i\alpha; 5/2; \chi) & = & \nonumber \\
	  & & \hspace{-14em} \frac{-3[\alpha\sinh \chi \cos(\alpha\chi) 
	- \cosh \chi \sin(\alpha\chi)]} {\alpha(1 + \alpha^2)\sinh^3 \chi } .
	\label{eq:hgdip2} 
	\end{eqnarray}
}	

\section{\label{app:Mannheim}Parameters in the model of Mannheim}

Mannheim finds that the field equations of conformal gravity take on a
particularly simple form in the context of a FRW space, as in cosmological
investigations. He obtains (\cite{mann6}, equation (230)) the following
expression for the expansion factor, $R$, as a function of the ordinary
time, $t$, assuming the matter in the Universe is in the form of radiation
and the space has negative curvature ($k=-1$):
\begin{equation}
R^2(t) = \frac{(\beta_{\mathrm{M}} - 1)}{2\alpha_{\mathrm{M}}} +
\frac{\beta_{\mathrm{M}} \sinh^2 (\alpha_{\mathrm{M}}^{1/2} ct)}
{\alpha_{\mathrm{M}}} .  \label{eq:manna}
\end{equation}
$\alpha_{\mathrm{M}}$ and $\beta_{\mathrm{M}}$ are real, positive
parameters. $\beta_{\mathrm{M}} $ is dimensionless and greater than unity;
$\alpha_{\mathrm{M}}$ has dimension ${\mathrm{length}}^{-2}$. (We write
$\alpha_{\mathrm{M}}$ rather than Mannheim's $\alpha$ to prevent any
possible confusion if $\alpha$ is used elsewhere, for example to denote
the fine structure constant; we use the same convention for other
parameters in Mannheim's model.) We will use (\ref{eq:manna}) for the whole
range of time from $t=0$ to the present, recognizing that this will be
inaccurate during the mass-dominated phase of the expansion. We estimate
this time interval below.

Time $t=0$ corresponds to the surface of the bubble. Here $R$ has a non-zero
minimum value,
$R_{\mathrm{min}} = [(\beta_{\mathrm{M}}-1)/(2\alpha_{\mathrm{M}})]^{1/2}$,
and a maximum temperature, $T_{\mathrm{max}}$. Mannheim defines a third
parameter, $A_{\mathrm{M}}$, by
\begin{equation}
A_{\mathrm{M}} = R^4 (t) T^4 (t) . \label{eq:A_Mdef}
\end{equation}
We will not try to explain the large size of $A_{\mathrm{M}}$, but will make
essential use of the fact that (in this simple model, at least) it has a
constant value during the expansion. If we can determine the three
parameters $\alpha_{\mathrm{M}}$, $\beta_{\mathrm{M}}$ and
$A_{\mathrm{M}}$ we can derive the values of all the rest of Mannheim's
parameters for this model.

Equation (\ref{eq:manna}) leads to an expression for the 
acceleration parameter, $q_0$. By comparing this with the value obtained
from supernova observations, Mannheim finds that
$\alpha_{\mathrm{M}}^{1/2} c t \approx 0.7$ at the present time, so that
$\alpha_{\mathrm{M}}^{1/2}$ is of the order of the inverse Hubble distance.
$A_{\mathrm{M}}$ can be estimated from the current CMB spectrum and the
Hubble distance; we will use a value of
$A_{\mathrm{M}}^{1/4} = 1.68 \times 10^{29}$.

The parameter $\beta_{\mathrm{M}}$ is harder to estimate. The temperature
evolution of the Universe is given by the following equation
(\cite{mann6}, equation (232)):
\begin{equation}
\frac{T_{\mathrm{max}}^2}{T^2} = 1 +
\frac{2\beta_{\mathrm{M}} \sinh^2 (\alpha_{\mathrm{M}}^{1/2} c t)}
{(\beta_{\mathrm{M}} - 1)} . \label{eq:tempev}
\end{equation}
Since $T_{\mathrm{max}} \gg T_0$, where $T_0$ is the current temperature,
$\beta_{\mathrm{M}}$ must be extremely close to $1$. It is therefore
convenient to define a new parameter,
$\delta_{\mathrm{M}} = \beta_{\mathrm{M}} - 1$, with
$0 < \delta_{\mathrm{M}} \ll 1$. Equation (\ref{eq:manna}) then shows
that $\delta_{\mathrm{M}}$ has an appreciable effect on $R(t)$ only at the
very earliest times. We will estimate $\delta_{\mathrm{M}}$ below.

Mannheim suggests that the CW phase transition occurs at an intermediate
temperature, $T_{\mathrm{V}}$, with
$T_{\mathrm{V}}^2 \approx T_{\mathrm{max}} T_0$. He regards the mechanism of
this phase transition as a matter for particle physics, and does not
consider it in detail. In our model the transition occurs at a considerably
higher temperature, and appears not to be connected to $T_{\mathrm{V}}$.

Define $x = \alpha^{1/2} t$, with $x_0 = \alpha^{1/2} t_0 = 0.7$ the current
value. The current CMB temperature is
$T_0 = 2.3 \times 10^{-4} {\mathrm{eV}}$.
The CW transition will occur at a temperature of about
$T_{\mathrm{cw}} = 10^9 {\mathrm{eV}}$, so,
since $\delta_{\mathrm{M}} \ll x_{\mathrm{cw}} \ll 1$:
\begin{eqnarray}
x_{\mathrm{cw}} & = & \sinh(0.7) (2.3 \times 10^{-4})/ 10^9 \nonumber \\
& = & 1.7 \times 10^{-13} , \label{eq:xCW} \\
\eta_0 - \eta_{\mathrm{cw}} & = & \int_{x_{\mathrm{cw}}}^{x_0} \!
\frac{\mathrm{d}x}{\sinh(x)} \nonumber \\
  & = & \ln \left(\frac{\tanh (x_0 /2)}{\tanh (x_{\mathrm{cw}} /2)} \right)
= 29 . \label{eq:e0eCW}
\end{eqnarray}

Also, since $x_{cw} \ll 1$,
\begin{eqnarray}
\eta_{\mathrm{cw}} & = & \int_0^{x_{\mathrm{cw}}} \!\!
\frac{\mathrm{d}x}{ \left[ \delta_{\mathrm{M}} /2 + x^2 \right]^{1/2}}
\nonumber \\
  & = & \ln \left[ \frac{ x + \sqrt{ \delta_{\mathrm{M}} /2 + x^2} }
{ \sqrt{\delta_{\mathrm{M}} /2}} \right] \nonumber \\
  & = & \ln \left( 2^{3/2} x_{\mathrm{cw}}
\delta_{\mathrm{M}}^{-1/2} \right) , \\
x_{\mathrm{cw}} & = &  2^{-3/2} \delta_{\mathrm{M}}^{1/2}
\exp (\eta_{\mathrm{cw}}) , \\
x_{\mathrm{min}} & = & 2^{-1/2} \delta_{\mathrm{M}}^{1/2} , \\
T_{\mathrm{max}} & = & T_{\mathrm{cw}} x_{\mathrm{cw}} / x_{\mathrm{min}}
\nonumber \\
  & = & 0.5 \times 10^9 \times \exp (\eta_{\mathrm{cw}})
= 3 \times 10^{13} \, {\mathrm{eV}} , \label{eq:Tmax}
\end{eqnarray}
where we have used $\eta_{\mathrm{cw}} = 11$, as derived from
(\ref{eq:p_exp_1}).

Equation (232) of \cite{mann6} gives
\ifthenelse {\lengthtest{\baselineskip > 16pt}}	
{
	\begin{eqnarray}
	\delta_{\mathrm{M}} & = & 2 (0.76)^2
	\left(\frac{ 2.3 \times 10^{-4}}{3 \times 10^{13}} \right)^2
	= 6.8 \times 10^{-35} , \label{eq:delta} \\
	T_{\mathrm{V}} & = &
	T_{\mathrm{max}} (\delta_{\mathrm{M}}/2)^{1/4} =
	7 \times 10^4 \, {\mathrm{eV}} .  \label{eq:TV}
	\end{eqnarray}
} {	
	\begin{eqnarray}
	\delta_{\mathrm{M}} & = & 2 (0.76)^2
	\left(\frac{ 2.3 \times 10^{-4}}{3 \times 10^{13}} \right)^2
	= 6.8 \times 10^{-35} , \qquad \label{eq:delta} \\
	T_{\mathrm{V}} & = &
	T_{\mathrm{max}} (\delta_{\mathrm{M}}/2)^{1/4} =
	7 \times 10^4 \, {\mathrm{eV}} .  \label{eq:TV}
	\end{eqnarray}
}	

Our equation (\ref{eq:manna}), Mannheim's expression for $R^2(t)$, will
become inaccurate during the mass-dominated phase of the expansion, between
$T=T_{\mathrm{cw}}$ and $T_{\mathrm{V}}$. This corresponds to an interval
of conformal time of about $10$, compared to $40$ for the whole interval
from $t=0$ to the present.

\section{\label{app:normalize}Normalization with conductivity}

In this appendix we will work in the usual spherical polar coordinates, and
make connection with Riemannian coordinates when necessary.

Suppose we have a dipole at the origin, oscillating with time dependence
$\exp (-i\omega t)$ in the $z$ direction. The surrounding medium is of
uniform conductivity, $\overline{\sigma}$, so that the current density,
${\mathbf j}$, is given by ${\mathbf j} = \overline{\sigma} {\mathbf E}$. We
will analyze this system by imagining a small sphere of radius $r_1$ cut out
of the medium surrounding the dipole. Induced currents flowing in the
medium will cause surface charges to appear on the sphere, and the total
dipole moment will be the sum of the original dipole moment and that
due to the induced charges. We assume the permittivity and magnetic
susceptibility are essentially unity, so ${\mathbf D} = {\mathbf E}$ and
${\mathbf B} = {\mathbf H}$. In such a system
$\nabla \cdot {\mathbf j} = 0$ follows from Maxwell's equations, so there
are no volume charges in the medium. 

The dipole moment at the center of the small sphere is denoted by 
$D_{\mathrm true} = D (\omega)\exp(-i\omega t)$, where $D (\omega)$ is
the true dipole strength at angular frequency $\omega$. The induced
dipole moment due to the surface charges is $D_{\mathrm ind}$, so the total
dipole moment is $D_{\mathrm tot} = D_{\mathrm true} + D_{\mathrm ind}$.

Just outside the sphere the electrostatic potential is given by
\begin{equation}
\Phi = D_{\mathrm tot}P_1 (\cos \theta )/r_1^2 ,
\label{eq:Phi_B}
\end{equation}
where $P_1 (\cos \theta ) = cos \theta$.
The radial component of ${\mathbf E}$ is given by 
\begin{equation}
E_r = 2D_{\mathrm tot} P_1 (\cos \theta )/r_1^3 .
\label{eq:E_r_B}
\end{equation}

The surface charge density, $s$, obeys the relation
\begin{equation}
\frac{ds}{dt} = -j_r \;,
\end{equation}
where $j_r$ is evaluated just outside the sphere. This gives
\begin{equation}
s = \frac{-i\overline{\sigma} E_r}{\omega} = s_0 P_1 (\cos \theta) ,
\end{equation}
where
\begin{equation}
s_0 = \frac{-2i\overline{\sigma} D_{\mathrm tot}}{r_1^3 \omega} \;.
\label{eq:s_0}
\end{equation}

The induced dipole moment, $D_{\mathrm ind}$, is then given by an integral
over the surface of the sphere:
\begin{eqnarray}
D_{\mathrm ind} & = & 2 \pi \int_{0}^{\pi} d \theta \sin \theta r_1^2 
[s_0 P_1 (\cos \theta)][r_1^2 P_1 (\cos \theta)] \nonumber \\
  & = & \left(\frac{-4\pi i \overline{\sigma}}{\omega}\right)
\left( \frac{2}{3} \right) D_{\mathrm tot} \, ,
\end{eqnarray}
giving
\begin{eqnarray}
D_{\mathrm tot} & = & D_{\mathrm true} + D_{\mathrm ind} \nonumber \\
  & = & \frac{3 \omega D_{\mathrm true}}
{3 \omega + 8\pi i \overline{\sigma}} \, .
\end{eqnarray}
 
We set $H_\phi = C_5 (\omega,\overline{\sigma}) N_1 (\theta) h(r)
\exp (-i \omega t)$, where $C_5 (\omega,\overline{\sigma})$ is the
normalizing function we are looking for, $N_1 = dP_1 /d\theta$, and
$h(r)$ satisfies
\begin{equation}
\frac{d^2 h}{d r^2} + \frac{2}{r} \frac{dh}{dr} - \frac{2h}{r^2}
+ \kappa^2h = 0 \;,
\end{equation}
with $\kappa^2 = \omega (\omega + 4\pi i \overline{\sigma})$.
We choose the solution that represents outgoing waves, so
\begin{eqnarray}
h(r) & = & h_1^{(1)} (\kappa r) \nonumber \\
  & = & \left( - \frac{i}{(\kappa r)^2}
- \frac{1}{\kappa r} \right) e^{i\kappa r} .
\end{eqnarray}
Here $h_1^{(1)} (\kappa r)$ is the spherical Bessel function defined in
AS, ch. 10.

The Maxwell equation 
$\nabla\times {\mathbf H} = -i(\omega + 4\pi i\overline{\sigma}){\mathbf E}$
then gives
\begin{equation}
E_r = \frac{-2iC_5 (\omega,\overline{\sigma}) P_1(\cos \theta) h_1^{(1)}
(\kappa r) \exp (-i \omega t)}{r (\omega + 4\pi i\overline{\sigma})} .
\end{equation}
For small $r$ this becomes:
\begin{equation}
E_r = \frac
{-2 C_5 (\omega,\overline{\sigma}) P_1(\cos \theta) \exp (-i \omega t)}
{\kappa^2 r^3 (\omega + 4\pi i \overline{\sigma})} .
\end{equation}

But also, for small $r$, we have (\ref{eq:E_r_B}), so
\ifthenelse {\lengthtest{\baselineskip > 16pt}}	
{
	\begin{eqnarray}
	C_5 (\omega,\overline{\sigma}) & = & \frac
	{-3 \omega \kappa^2 (\omega + 4\pi i \overline{\sigma}) D (\omega) }
	{(3\omega + 8\pi i \overline{\sigma})} , \label{eq:C3norm} \\
	H_{\phi} & = & \frac
	{-3\omega \kappa^2 (\omega + 4\pi i \overline{\sigma}) D (\omega)
	N_1 (\theta) h_1^{(1)} (\kappa r) \exp (-i \omega t)}
	{(3\omega + 8\pi i \overline{\sigma})} , \label{eq:H_B} \\
	E_r & = & \frac
	{6 i \omega \kappa^2 D (\omega) P_1 (\theta) h_1^{(1)} (\kappa r)
	\exp (-i \omega t)} { r (3\omega + 8 \pi i \overline{\sigma})} .
	\label{eq:E_B}
	\end{eqnarray}
} {	
	\begin{eqnarray}
	C_5 (\omega,\overline{\sigma}) & = & \frac
	{-3 \omega \kappa^2 (\omega + 4\pi i \overline{\sigma}) D (\omega) }
	{(3\omega + 8\pi i \overline{\sigma})} , \label{eq:C3norm} \\
	H_{\phi} & = & \frac
	{-3\omega \kappa^2 (\omega + 4\pi i \overline{\sigma}) D (\omega)}
	{(3\omega + 8\pi i \overline{\sigma})} \label{eq:H_B} \nonumber \\
	& & \times N_1 (\theta) h_1^{(1)} (\kappa r) \exp (-i \omega t) , \\
	E_r & = & \frac
	{6 i \omega \kappa^2 D (\omega) P_1 (\theta) h_1^{(1)} (\kappa r)
	\exp (-i \omega t)} { r (3\omega + 8 \pi i \overline{\sigma})} .
	\qquad \quad \label{eq:E_B}
	\end{eqnarray}
}	

In this appendix we have used ordinary polar coordinates in flat space.
We need now to transform to coordinates of a FRW flat space, with metric
\ifthenelse {\lengthtest{\baselineskip > 16pt}}	
{
	\begin{eqnarray}
	\mathrm{d}s^2 & = & R^2 (\eta)\left(-\mathrm{d} \eta^2 +
	\mathrm{d}\chi^2 + \chi^2 \, \mathrm{d}\theta^2
	+ \chi^2 \sin^2 \theta \, \mathrm{d}\phi^2 \right) ,
	\label{eq:ds4}
	\end{eqnarray}
} {	
	\begin{eqnarray}
	\mathrm{d}s^2 & = & R^2 (\eta)\left(-\mathrm{d} \eta^2 +
	\mathrm{d}\chi^2 + \chi^2 \, \mathrm{d}\theta^2 \right. \nonumber \\
	  & & \left. {} + \chi^2 \sin^2 \theta \, \mathrm{d}\phi^2 \right) ,
	\label{eq:ds4}
	\end{eqnarray}
}	

To get from $H_{\phi}$ of (\ref{eq:H_B}) to $F_{12}$ of (\ref{eq:F12def}),
in the flat FRW metric (\ref{eq:ds4}), we first follow Weinberg \cite{wein2},
ch. 4, sec. 8, and multiply by $\chi$. We then convert from $t$ to $\eta$
and $r$ to $\chi$ by multiplying by $R^2 (\eta)$. We also convert
$\omega$ to $n$, $\kappa$ to $k$ and $\overline{\sigma}$ to
$\sigma = R(\eta) \overline{\sigma}$:
\ifthenelse {\lengthtest{\baselineskip > 16pt}}	
{
	\begin{equation}
	F_{12} = \frac
	{-3 n k^2 (n + 4\pi i \sigma) D (n) N_1 (\theta) \chi h_1^{(1)} (k \chi)
	\exp (-i n \eta)}{R(\eta) (3n + 8 \pi i \sigma)} .
	\label{eq:F12flat}
	\end{equation}
} {	
	\begin{eqnarray}
	F_{12} & = & \frac
	{-3 n k^2 (n + 4\pi i \sigma) D (n)}{R(\eta) (3n + 8 \pi i \sigma)}
	\nonumber \\
	  & & \times N_1 (\theta) \chi h_1^{(1)} (k \chi) \exp (-i n \eta) .
	\label{eq:F12flat}
	\end{eqnarray}
}	

We can obtain the corresponding function in curved space by 
multiplying (\ref{eq:f3out2perm}) by $N_1 (\theta) \exp(-in \eta)$:
\ifthenelse {\lengthtest{\baselineskip > 16pt}}	
{
	\begin{equation}
	F_{12} = \frac{C_3 (n,\sigma) N_1 (\theta) \exp [i(k \chi - n \eta)]}
	{2 u} \left(1 - 2iku + u^2 \right) . \label{eq:F12curv}
	\end{equation}
} {	
	\begin{eqnarray}
	F_{12} & = & \frac{C_3 (n,\sigma) N_1 (\theta)
	\exp [i(k \chi - n \eta)]} {2 u} \nonumber \\
	  & & \times \left(1 - 2iku + u^2 \right) . \label{eq:F12curv}
	\end{eqnarray}
}	

We can now find $C_3 (n,\sigma)$ by matching (\ref{eq:F12flat}) with 
(\ref{eq:F12curv}) for small $\chi$:
\begin{equation}
C_3 (n,\sigma) = \frac{3in (n + 4\pi i \sigma) D (n)}
{R(\eta) (3n + 8 \pi i \sigma)} .
\label{eq:normC3perm}
\end{equation}

With $D (n)$ given by (\ref{eq:dipstrength}), we can check that in the limit
$\sigma \rightarrow 0$, (\ref{eq:normC3perm}) tends to
(\ref{eq:normC1}), as it should.

\section{\label{app:energy_transfer}Derivation of the energy-transfer
equation}

In this appendix we present details of the derivation of the energy-transfer
equation (\ref{eq:econ6}), starting from equation (\ref{eq:econ2}):
\ifthenelse {\lengthtest{\baselineskip > 16pt}}	
{
	\begin{eqnarray}
	\sqrt{g}\, \mathrm{d}^4 x\, U_0 \left[ \frac{1}{\sqrt{g}}
	\frac{\partial}{\partial x^0 } \left( \sqrt{g}\, T^{00} \right) 
	+ \frac{1}{\sqrt{g}} \frac{\partial}{\partial x^i } 
	\left( \sqrt{g}\, T^{i 0} \right) + \Gamma^{0}_{ii} T^{ii}
	+ \Gamma^{0}_{00} T^{00} \right] = 0 . \label{eq:econ2_app}
	\end{eqnarray}
} {	
	\begin{eqnarray}
	{} & &\sqrt{g}\, \mathrm{d}^4 x\, U_0 \left[ \frac{1}{\sqrt{g}}
	\frac{\partial}{\partial x^0 } \left( \sqrt{g}\, T^{00} \right) 
	+ \frac{1}{\sqrt{g}} \frac{\partial}{\partial x^i } 
	\left( \sqrt{g}\, T^{i 0} \right) \right. \nonumber \\
	{} & & \left. {} + \Gamma^{0}_{ii} T^{ii}
	+ \Gamma^{0}_{00} T^{00} \right] = 0 . \label{eq:econ2_app}
	\end{eqnarray}
}	

Write this as 
\begin{equation}
z_1 + z_2 + z_3 + z_4 = 0, \label{eq:zi}
\end{equation}
where the $z_i$ are derived from the four terms in the bracket in
(\ref{eq:econ2_app}). The strategy will be to get an expression for
$z_2$ from the propagation of $\mathbf{E}$ and $\mathbf{H}$. We combine
this with $z_1$, $z_3$ and $z_4$, and arrive at an integrodifferential
equation for  $T^{00}$.

\subsection{\label{subsec:z1_app}Calculation of $\bm{z_1}$}

We consider first the effect of a single source particle at the origin of
coordinates, and let the duration of our 4-volume be $\Delta x^0$.
Our coordinate volume $\mathrm{d} \chi \mathrm{d} \theta \mathrm{d} \phi$
is a small part of a spherical shell of radius $\chi$ surrounding the source
particle. Then $z_1$ in (\ref{eq:zi}) is given by
\begin{eqnarray}
z_1 & = & \Delta x^0 U_0 \frac{\partial}{\partial x^0 } 
\left(\sqrt{g} \, T^{00} \right) \mathrm{d} \chi \mathrm{d} \theta
\mathrm{d} \phi \nonumber \\
  & = & R^4 \left( 4 \dot{R} T^{00} + R \dot{T}^{00} \right)
\sinh^2 \chi \sin \theta
\Delta \eta \mathrm{d} \chi \mathrm{d} \theta \mathrm{d} \phi . \qquad 
\label{eq:z1_1}
\end{eqnarray}

\subsection{\label{subsec:z2_app}Calculation of $\bm{z_2}$}

For $z_2$ in (\ref{eq:zi}) the only relevant derivative is
$\partial / \partial \chi$, and we get
\begin{eqnarray}
z_2 & = & \Delta x^0 \mathrm{d} \theta \, \mathrm{d} \phi \, U_0
\Delta \left( \sqrt{g} \, T^{10} \right) \nonumber \\
  & = & \Delta \eta R \frac{\partial}{\partial \chi } \left( \sqrt{g} T^{10}
\right) \mathrm{d} \chi \mathrm{d} \theta \mathrm{d} \phi .
\label{eq:z2_1}
\end{eqnarray}
Here $\Delta (\sqrt{g} \, T^{10} )$  is the net outward flux from the shell,
for a pulse that protrudes on both sides. The pulse is, of course,
attenuated on the outward side, so the total outward flux is negative.

\subsection{\label{subsec:z34_app}Calculation of $\bm{z_3}$ and $\bm{z_4}$}

To compute $z_3$ and $z_4$ in (\ref{eq:zi}) we start from Weinberg (5.4.2):
\begin{equation}
T^{\mu\nu} = p g^{\mu\nu} + ( p + \rho ) U^{\mu} U^{\nu} .
\label{eq:wein5_4_2}
\end{equation}

The only Christoffel symbols we need are $\Gamma^{0}_{00} = \dot{R}/R$
and $\Gamma^{0}_{ij} = \dot{R} g_{ij} / R^3$, where a dot denotes
$\partial / \partial \eta$, and $i,\, j$ take values 1, 2, 3. Then
\begin{eqnarray}
T^{00} & = & \frac{-p}{R^2} + \frac{p+ \rho}{R^2} = \frac{\rho}{R^2} ,
\label{eq:T00_1} \\
T^{ij} & = & p g^{ij} , \\
\Gamma^0_{\mu\lambda} T^{\mu\lambda} & = & \frac{ \dot{R} \rho}{R^3}
+ \frac{ \dot{R} p \, g_{ij} \, g^{ij}}{R^3} \nonumber \\
  & = & \frac{ 2 \rho \dot{R}}{R^3} \quad \mathrm{if} \quad p = \rho/3 .
\end{eqnarray}

We now have
\begin{eqnarray}
z_3 + z_4 & = & U_0 \frac{2 \rho \dot{R}}{R^3} \Delta x^0 \mathrm{d}\chi\,
\mathrm{d}\theta\,\mathrm{d}\phi\, \sqrt{g} \nonumber \\
  & = & 2 T^{00} \dot{R} \Delta \eta \sqrt{g} \mathrm{d} \chi 
\mathrm{d} \theta \mathrm{d} \phi . \label{eq:z34_1}
\end{eqnarray}

\subsection{\label{subsec:energytransfer_T}Energy-transfer equation in
terms of $\bm{T^{\mu \nu}}$}

Using (\ref{eq:z1_1}), (\ref{eq:z2_1}) and (\ref{eq:z34_1})
in (\ref{eq:zi}) we get
\ifthenelse {\lengthtest{\baselineskip > 16pt}}	
{
	\begin{eqnarray}
	4 \dot{R} T^{00} + R \dot{T}^{00} +
	\frac{R}{\sinh^2 \chi} \frac{\partial}{\partial \chi}
	\left( T^{10} \sinh^2 \chi\right) + 2 \dot{R} T^{00} & = & 0 ,
	\nonumber \\
	\frac{6 \dot{R} T^{00}}{R} + \dot{T}^{00} +
	\frac{1}{\sinh^2 \chi} \frac{\partial}{\partial \chi}
	\left( T^{10} \sinh^2 \chi \right) & = & 0 .
	\label{eq:econ3}
	\end{eqnarray}
} {	
	\begin{eqnarray}
	4 \dot{R} T^{00} + R \dot{T}^{00} +
	\frac{R}{\sinh^2 \chi} \frac{\partial}{\partial \chi}
	\left( T^{10} \sinh^2\chi\right) & & \nonumber \\
	{} + 2 \dot{R} T^{00} & = & 0 , \nonumber \\
	\frac{6 \dot{R} T^{00}}{R} + \dot{T}^{00} +
	\frac{1}{\sinh^2 \chi} \frac{\partial}{\partial \chi}
	\left( T^{10} \sinh^2 \chi \right) & = & 0 . \qquad
	\label{eq:econ3}
	\end{eqnarray}
}	

We must now develop this equation in two ways:
\begin{enumerate}
\item Express the first two terms as functions of the temperature,
$T(\eta)$.
\item Sum over all the source charges that contribute to $T^{10}$.
Since we are concerned here with the main pulse, not the tail, this implies
an integration along the light cone.
\end{enumerate}

\subsection{\label{subsec:TandT}Expression of $\bm{T^{00}}$ in terms of
$\bm{T(\eta)}$}

We will measure temperature in electron-volts, $R$ in meters, and will
assume the energy density is given in terms of the temperature by the
familiar formulae for electromagnetic radiation (we will clearly miss a
numerical factor here when dealing with the real plasma, but the overall
relationship will be correct). Denote by $q_{\mathrm{MKS}}$ the elementary
charge in MKS units, and let the energy density be $\rho$. Then
\begin{eqnarray}
\rho & = & P_1 T^4 \,\mathrm{J \cdot m^{-3}} , \nonumber \\
P_1 & = & \frac{\pi^2}{15} q_{\mathrm{MKS}}^4
\cdot \hbar^{-3} \cdot c^{-3} \nonumber \\
  & = & \frac{\pi^2 \left(1.6 \times 10^{-19}\right)^4}
{15 \left(1.05 \times 10^{-34}\right)^3 \left(3 \times 10^8\right)^3} \\
  & = & 13.8 \, \mathrm{\left(eV\right)^{-4} \cdot J \cdot m^{-3}} ,
\label{eq:P1def} \\
T^{00} & = & \frac{P_1 T^4 (\eta ) }{ R^2(\eta ) }
\,\mathrm{J \cdot m^{-5}} , \label{eq:T00_2} \\
\dot{T}^{00} & = & \frac{4 P_1 T^3 (\eta ) \dot{T} (\eta )}{ R^2(\eta ) }
- \frac{2 P_1 T^4 (\eta ) \dot{R} (\eta )}{ R^3 (\eta )}
\,\mathrm{J \cdot m^{-5}} . \qquad \label{eq:Tdot00_2}
\end{eqnarray}

\subsection{\label{subsec:lightcone}Integrating along the light cone}

Using the asymptotic forms (\ref{eq:F12asymptotic}) and
(\ref{eq:F20asymptotic}) for the fields, we obtain
\ifthenelse {\lengthtest{\baselineskip > 16pt}}	
{
	\begin{eqnarray}
	T^{10} & = & - g^{00} g^{11} g^{22} F_{12} F_{20} \nonumber \\
	  & = &  \frac{(3 q V \sin \theta )^2}
	{R^6 (\eta ) \sinh^2 \chi } \frac{1}{4 \pi (\eta + \tau )}
	\exp \left( \frac{-2 \pi \sigma \chi^2 }{\eta + \tau} \right) , \;
	\label{eq:T10} \\
	\frac{\partial}{\partial \chi }\left( T^{10} \sinh^2 \chi \right) & = &
	\frac{-(3 q V \sin \theta )^2 \sigma \chi}
	{R^6 (\eta ) (\eta + \tau)^2}
	\exp \left( \frac{-2 \pi \sigma \chi^2 }{\eta + \tau} \right) . 
	\label{eq:T10deriv}
	\end{eqnarray}
} {	
	\begin{eqnarray}
	T^{10} & = & - g^{00} g^{11} g^{22} F_{12} F_{20} \nonumber \\
	  & = &  \frac{(3 q V \sin \theta )^2}
	{R^6 (\eta ) \sinh^2 \chi } \frac{1}{4 \pi (\eta + \tau )}
	\exp \left( \frac{-2 \pi \sigma \chi^2 }{\eta + \tau} \right) ,
	\quad \quad \; \label{eq:T10} 
	\end{eqnarray}
	\vspace{-1em}
	\begin{eqnarray}
	\frac{\partial}{\partial \chi }\left( T^{10} \sinh^2 \chi \right) & = &
	\nonumber \\
	& & \hspace{-4em} \frac{-(3 q V \sin \theta )^2 \sigma \chi}
	{R^6 (\eta ) (\eta + \tau)^2} 
	\exp \left( \frac{-2 \pi \sigma \chi^2 }{\eta + \tau} \right) .
	\quad \label{eq:T10deriv}
	\end{eqnarray}
}	

This form refers to a single source particle. To get the total $T^{10}$
term in (\ref{eq:econ3}) we must integrate along the light cone.
We will need the number density, which we write as $P_2 T^3$, with the
constant $P_2$ defined as for electromagnetic radiation:
\ifthenelse {\lengthtest{\baselineskip > 16pt}}	
{
	\begin{eqnarray}
	P_2 & = & \frac{2}{\pi^2} \zeta(3) \frac{q_{\mathrm{MKS}}^3}
	{\hbar^3 c^3} \nonumber \\
	  & = & 0.24 \left[\frac{1.6 \times 10^{-19}}
	{\left(1.05 \times 10^{-34} \right) \left(3 \times 10^8 \right)}
	\right]^3 = 3.15 \times 10^{20} \,
	\mathrm{\left(eV\right)^{-3} \cdot m^{-3}} . \label{eq:P2def}
	\end{eqnarray}
} {	
	\begin{eqnarray}
	P_2 & = & \frac{2}{\pi^2} \zeta(3) \frac{q_{\mathrm{MKS}}^3}
	{\hbar^3 c^3} \nonumber \\
	  & = & 0.24 \left[\frac{1.6 \times 10^{-19}}
	{\left(1.05 \times 10^{-34} \right) \left(3 \times 10^8 \right)}
	\right]^3 \nonumber \\
	& = & 3.15 \times 10^{20} \,
	\mathrm{\left(eV\right)^{-3} \cdot m^{-3}} . \label{eq:P2def}
	\end{eqnarray}
}	

We can now use (\ref{eq:T10deriv}) to write an expression for the total
$T^{10}$ term in (\ref{eq:econ3}). We neglect $\tau$ in comparison with
$\eta$, and write $\eta - \eta^{\,\prime}$ in place of $\eta$. We also
express the charge $q$, which is in Gaussian units, in terms of
$q_{\mathrm{MKS}}$:
\begin{equation}
q = \frac{1}{\sqrt{4 \pi \epsilon_0}} q_{\mathrm{MKS}} \, ,
\end{equation}
with $\epsilon_0 = (1/36\pi ) \times 10^{-9} \,
\mathrm{C^2 \cdot J^{-1} \cdot m^{-1}}$.
\ifthenelse {\lengthtest{\baselineskip > 16pt}}	
{
} {	
	\begin{widetext}
}	
\begin{eqnarray}
\left[ \frac{1}{\sinh^2 \chi} \frac{\partial }{\partial \chi}
\left( T^{10} \sinh^2 \chi \right) \right] & = &
\frac{-1}{R^6 (\eta ) } \int_0^{\eta} \mathrm{d} \eta^{\,\prime}
\int_0^{\infty} \mathrm{d} \chi \delta (\chi - \eta + \eta^{\,\prime} )
\left[ P_2 T^3 (\eta^{\,\prime}) \right] \nonumber \\
  & & \hspace{-12em} \times \left(4 \pi R^3(\eta^{\,\prime}) \sinh^2 \chi
\right) \left(\frac{1}{\sinh^2 \chi}\right)
\frac{(3 q_{\mathrm{MKS}})^2 \langle V^2 \rangle
\langle \sin^2 \theta \rangle \sigma \chi}
{4 \pi \epsilon_0 (\eta - \eta^{\,\prime})^2}
\exp \left( \frac{-2 \pi \sigma \chi^2 }{\eta - \eta^{\,\prime}} \right)
\Delta \eta \, . \quad 
\label{eq:z2tot1}
\end{eqnarray}

As in section \ref{sec:mass_1} we set
$\langle V^2 \rangle \approx 1/(2 + p(\eta ))$ and
$\langle \sin^2 \theta \rangle = 2/3$; we also define a new constant $P_3$:
\ifthenelse {\lengthtest{\baselineskip > 16pt}}	
{
	\begin{eqnarray}
	P_3 & = & P_2 \times 36\pi \times \left(2/3 \right) \times
	\left( 1.6 \times 10^{-19} \right)^2 \times
	\left( 9 \times 10^9 \right) \nonumber \\
	  & = & 5.47 \times 10^{-6} \,
	\mathrm{\left(eV\right)^{-3} \cdot J \cdot m^{-2}} ,
	\label{eq:P3def} \\
	\left[ \frac{1}{\sinh^2 \chi} \frac{\partial }{\partial \chi}
	\left( T^{10} \sinh^2 \chi \right) \right]
	& = & \nonumber \\
	& & \hspace{-8em} \frac{-P_3}{R^6 (\eta ) }
	\int_0^{\eta} \mathrm{d} \eta^{\,\prime}
	\frac{T^3 (\eta^{\,\prime}) R^3 (\eta^{\,\prime})}{2 + p(\eta)}
	\sigma (\eta - \eta^{\,\prime})
	\exp \left[ -2 \pi \sigma (\eta - \eta^{\,\prime}) \right] \Delta \eta 
	\, .  \label{eq:z2tot2}
	\end{eqnarray}
} {	
	\begin{eqnarray}
	P_3 & = & P_2 \times 36\pi \times \left(2/3 \right) \times
	\left( 1.6 \times 10^{-19} \right)^2 \times
	\left( 9 \times 10^9 \right) \nonumber \\
	  & = & 5.47 \times 10^{-6} \,
	\mathrm{\left(eV\right)^{-3} \cdot J \cdot m^{-2}} ,
	\label{eq:P3def} \\
	\left[ \frac{1}{\sinh^2 \chi} \frac{\partial }{\partial \chi}
	\left( T^{10} \sinh^2 \chi \right) \right]
	& = & \frac{-P_3}{R^6 (\eta ) } \int_0^{\eta} \mathrm{d} \eta^{\,\prime}
	\frac{T^3 (\eta^{\,\prime}) R^3 (\eta^{\,\prime})}{2 + p(\eta)}
	\sigma (\eta - \eta^{\,\prime})
	\exp \left[ -2 \pi \sigma (\eta - \eta^{\,\prime}) \right] \Delta \eta 
	\, .  \label{eq:z2tot2}
	\end{eqnarray}
}	

\ifthenelse {\lengthtest{\baselineskip > 16pt}}	
{
} {	
	\end{widetext}
}	

\subsection{\label{subsec:finalform}Final form of the energy-transfer
equation}

Using (\ref{eq:T00_2}), (\ref{eq:Tdot00_2}) and (\ref{eq:z2tot2})
in (\ref{eq:econ3}):
\ifthenelse {\lengthtest{\baselineskip > 16pt}}	
{
	\begin{eqnarray}
	{} & & 4 P_1 T^3 (\eta ) \dot{T} (\eta ) R^4 (\eta )
	+ 4 P_1 T^4 (\eta ) R^3 (\eta ) \dot{R} (\eta ) \nonumber \\
	{} & & {} - P_3 \int_0^{\eta} \mathrm{d} \eta^{\,\prime}
	\frac{T^3 (\eta^{\,\prime}) R^3 (\eta^{\,\prime})}{2 + p(\eta )}
	\sigma (\eta - \eta^{\,\prime})
	\exp \left[ -2 \pi \sigma (\eta - \eta^{\,\prime}) \right] = 0 .
	\label{eq:econ5}
	\end{eqnarray}
} {	
	\begin{eqnarray}
	{} & & 4 P_1 T^3 (\eta ) \dot{T} (\eta ) R^4 (\eta )
	+ 4 P_1 T^4 (\eta ) R^3 (\eta ) \dot{R} (\eta ) \nonumber \\
	{} & & {} - P_3 \int_0^{\eta} \mathrm{d} \eta^{\,\prime}
	\frac{T^3 (\eta^{\,\prime}) R^3 (\eta^{\,\prime})}{2 + p(\eta )}
	\sigma (\eta - \eta^{\,\prime}) \nonumber \\
	{} & & \times \exp \left[ -2 \pi \sigma (\eta - \eta^{\,\prime})
	\right] = 0 . \label{eq:econ5}
	\end{eqnarray}
}	

This equation can be written in terms of the single variable
$A_{\mathrm{M}} (\eta ) \equiv T^4 (\eta ) R^4 (\eta )\,
\mathrm{eV^4 \cdot m^4}$, and a new constant, $P_4$:
\ifthenelse {\lengthtest{\baselineskip > 16pt}}	
{
	\begin{eqnarray}
	{} & & \dot{A}_{\mathrm{M}} (\eta ) -
	P_4 \int_0^{\eta} \mathrm{d} \eta^{\,\prime}
	A_{\mathrm{M}}^{3/4} (\eta^{\,\prime})
	\frac{\sigma (\eta - \eta^{\,\prime})}{2 + p(\eta )}
	\exp \left[ -2 \pi \sigma (\eta - \eta^{\,\prime}) \right] = 0 \, ,
	\label{eq:econ6_app} \\
	{} & & P_4 = P_3/P_1 = 3.97 \times 10^{-7}\,\mathrm{eV \cdot m} \, .
	\label{eq:P4def_app} 
	\end{eqnarray}
} {	
	\begin{eqnarray}
	{} & & \dot{A}_{\mathrm{M}} (\eta ) -
	P_4 \int_0^{\eta} \mathrm{d} \eta^{\,\prime}
	A_{\mathrm{M}}^{3/4} (\eta^{\,\prime})
	\frac{\sigma (\eta - \eta^{\,\prime})}{2 + p(\eta )}
	\nonumber \\
	{} & & \times \exp \left[ -2 \pi \sigma (\eta - \eta^{\,\prime})
	\right] = 0 \, , \label{eq:econ6_app} \\
	{} & & P_4 = P_3/P_1 = 3.97 \times 10^{-7}\,\mathrm{eV \cdot m} \, .
	\label{eq:P4def_app} 
	\end{eqnarray}
}	



\end{document}